\documentclass[showpacs,showkeys,preprintnumbers,superscriptaddress,amsmath,floatfix,amssymb,a4paper,secnumarabic]{revtex4}
\usepackage[colorlinks=true]{hyperref}
\usepackage{graphicx}
\usepackage{slashbox}

\newcommand{\beq}{\begin{equation}}
\newcommand{\eeq}{\end{equation}}

\def\tr{\hbox{Tr}}
\def\dalam{\hbox
{\vrule\vbox{\hrule\hbox to 1ex{ \hfill}\kern 1 ex\hrule}\vrule}}

\def\1/2{\hbox{$ {1 \over 2}$ }}

\def\arctg{\hbox{arctg}}

\def\inf{\infty}
\def\pd{\partial}

\def\a{\alpha} 
\def\b{\beta} 
 
\def\g{\gamma} \def\tg{\tilde {\g}}\def\G{\Gamma} 
\def\d{\delta} \def\D{\Delta}
\def\l{\lambda} 
\def\e{\epsilon}

\def\s{\sigma}
\def\r{\rho} 
\def\x{\xi}
\def\c{\chi} 
\def\vf{\varphi} 
\def\f{\phi} \def\F{\Phi}
 
\def\p{\psi} \def\P{\Psi}

\def\m{\mu}
\def\n{\nu}

\def\<{\langle}
\def\>{\rangle}

\def\({\left(}
\def\[{\left[}
\def\){\right)}
\def\]{\right]}

\begin{document}
\sloppy

\title{Vacuum  energy of one-dimensional supercritical Dirac-Coulomb system}
\author{A.~Davydov}
\email{davydov.andrey@physics.msu.ru} \affiliation{Department of Physics and
Institute of Theoretical Problems of MicroWorld, Moscow State
University, 119991, Leninsky Gory, Moscow, Russia}

\author{K.~Sveshnikov}
\email{costa@bog.msu.ru} \affiliation{Department of Physics and
Institute of Theoretical Problems of MicroWorld, Moscow State
University, 119991, Leninsky Gory, Moscow, Russia}

\author{Yu.~Voronina}
\email{voroninayu@physics.msu.ru} \affiliation{Department of Physics and
Institute of Theoretical Problems of MicroWorld, Moscow State
University, 119991, Leninsky Gory, Moscow, Russia}

\begin{abstract}
Nonperturbative vacuum polarization effects   are explored for  a supercritical Coulomb source with $Z>Z_{cr}$ in 1+1 D. Both the  vacuum charge density $\r_{vac}(x)$ and vacuum energy $E_{vac}$ are considered. It is shown that in the overcritical region  the behavior of vacuum energy could be significantly different from perturbative quadratic growth up to decrease reaching large negative values.
\end{abstract}


\maketitle

\section{Introduction}

Starting from Elliott and  Loudon \cite{1,2}, there is a lot of interest to the study of quasi-one-dimensional systems with Coulomb interaction, caused by the continuous growth of various physical applications \cite{3,4,5,6,7,8}. In this paper  we explore  the main nonperturbative features of a one-dimensional supercritical Dirac-Coulomb (DC) system. Whereas there is a lot of work devoted to such systems in 3+1 D (see, e.g., Refs.~\cite{9}\nocite{10,11,12}-\cite{13} and references therein), their 1+1 D analog has not been studied at all. Meanwhile, in Refs.~\cite{14}\nocite{15,16,17,18,19}-\cite{20}, it was shown that in superstrong  homogeneous magnetic fields the effective relativistic dynamics of the electronic component  in H-like atoms turns out to be quasi-one-dimensional, while the first critical charge $Z_{cr,1}$ could be less than $\simeq 170$ in absence of the field \cite{17,18,19,20}.

A separate attention is drawn to nonperturbative vacuum polarization effects, caused by diving of discrete levels into lower continuum in supercritical static or adiabatically slow varying Coulomb fields \cite{9,10,11,12,13}.  This work explores such essentially nonperturbative vacuum effects for a model of supercritical DC system in one-dimensional case,  with the main attention drawn to the vacuum polarization energy $E_{vac}$. Although the most of works consider the vacuum charge density $\r_{vac}(x)$ as the main polarization observable, by means of which, in particular, the contribution of vacuum polarization to the Lamb shift is calculated, $E_{vac}$  turns out to be not less informative and in many respects complementary to $\r_{vac}(x)$. Moreover, compared to $\r_{vac}(x)$, the main nonperturbative effects, which appear in vacuum polarization for $Z > Z_{cr,1}$ due to levels diving into lower continuum, show up in the behavior of vacuum energy even more clear, demonstrating explicitly their possible role in the  supercritical region.

For these purposes we consider here a simplified semi-analytic model with the external Coulomb source regulated via smooth cutoff at the scale  $a>0$  \nocite{21,22,23,24,25,26}\cite{17,27}
\beq
\label{0.1}
V(x)= - \frac{Z \a}{|x|+a} \ ,
\eeq
which allows to perform most part of calculations in analytical form, while the resulting qualitative picture turns out to be quite general.

As in other works on vacuum polarization in the strong  Coulomb field \cite{21,22,23,24}, radiative corrections from virtual photons are neglected. Henceforth, if it is not stipulated separately, relativistic units  $\hbar=m_e=c=1$ are used. Thence the electromagnetic coupling constant $\a=e^2$ is also dimensionless,  and numerical calculations, illustrating the general picture, are performed for  $\a=1/137.036$.

\section{Vacuum Energy in Perturbative Approach for 1+1 D}

\begin{figure}[h!]
\centerline{\includegraphics[width=50mm]{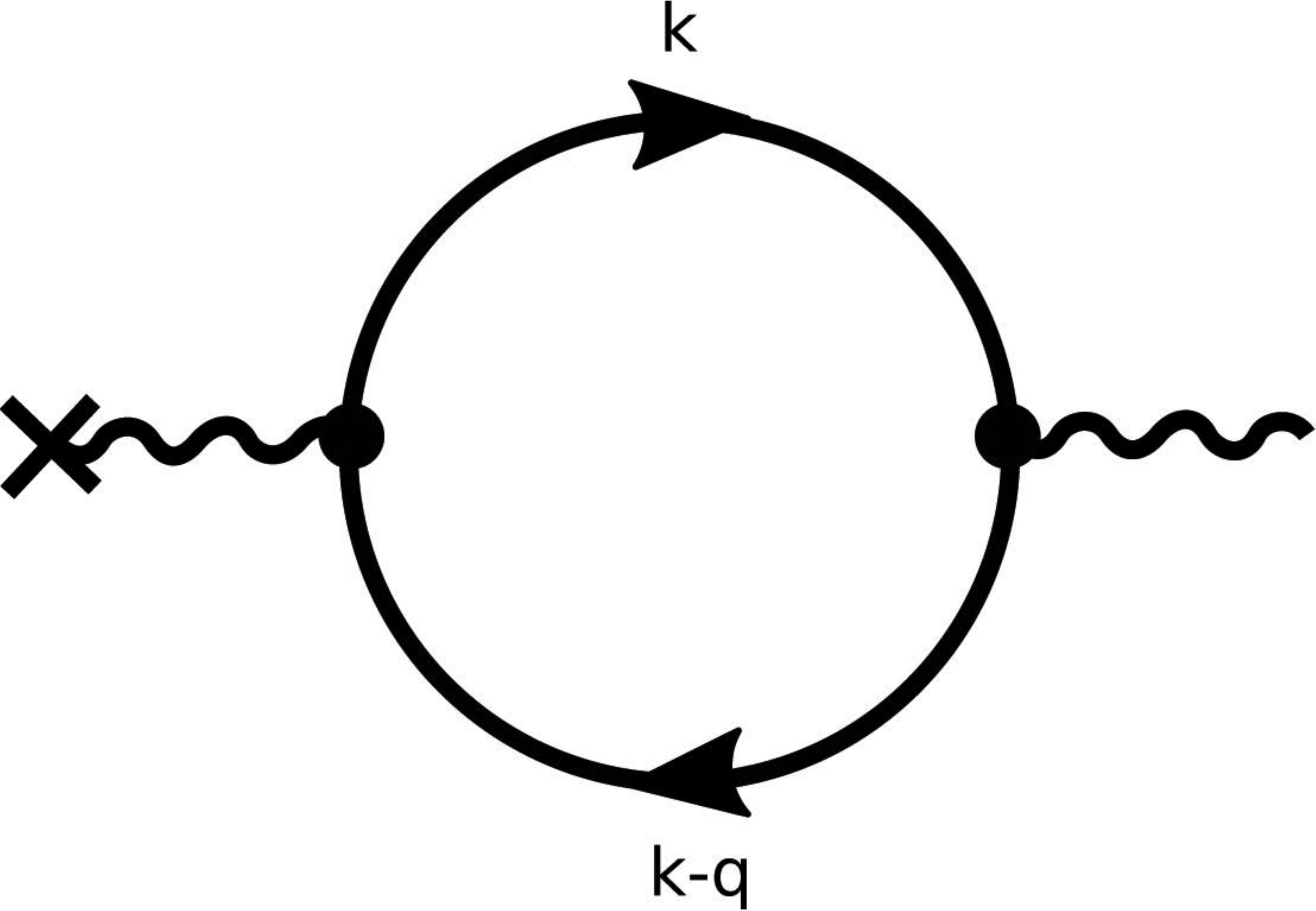}}
\caption{The lowest-order Feynman graph for vacuum polarization in the external field. \label{fig1}}
\end{figure}

In 1+1 D the expression for the regularized polarization operator $\Pi_{\m \n}^R(q)$, corresponding to the Feynman graph on Fig.~\ref{fig1}, takes the form
\beq
\label{1.1}
\begin{aligned}
&\Pi_{\m\n}^{R}(q)=\(q_{\m}q_{\n}- g_{\m\n} q^2\)\Pi^{R}(q^2)\ , \\
&\Pi^{R}(q^2)=\frac{4 \a}{m^2} \int\limits_0^{1} d\b \, \b(1-\b)\[ 1 -  \b(1-\b) \frac{q^2}{m^2-i \varepsilon}\]^{-1}.
\end{aligned}
\eeq
\\
In particular, for $q^2=-q_x^2$
\beq
\label{1.2}
\Pi^{R}(-q_x^2)=\frac{4 \a}{q_x^2}\(1-\frac{4 m^2}{q_x\sqrt{4 m^2+q_x^2}}\mathrm{arcsinh}\(\frac{q_x}{2m}\) \).
\eeq
In the perturbative approach, the vacuum polarization energy to the leading order is given by
\beq
\label{1.3}
E^{(1)}_{vac}=\frac{1}{2} \int\limits_{-\infty}^{+\infty}d x\, \r^{(1)}_{vac}(x)A_{0}^{ext}(x),
\eeq
where $A_{0}^{ext}$ is the external Coulomb source potential, which is assumed to be static, while  $\r^{(1)}_{vac}$ is the vacuum density, determined from the polarization potential
\beq
\label{1.4}
\r^{(1)}_{vac}(x)=-\frac{1}{4 \pi} \frac{d^2}{d x^2}\, A^{(1)}_{vac,0}(x).
\eeq
The one-loop (Uehling) vacuum polarization potential $A^{(1)}_{vac,0}$ is found from  $A_0^{ext}$ by means of the  polarization function $\Pi^{R}(-q_x^2)$  \cite{25}, namely
\beq
\label{1.5}
A^{(1)}_{vac,0}(x)=\frac{1}{2 \pi} \int\limits_{-\infty}^{+\infty}d q\,\mathrm{e}^{i q x} \Pi^{R}(-q^2)\widetilde{A}_{0}(q), \quad
\widetilde{A}_{0}(q)=\int\limits_{-\infty}^{+\infty}d y\,\mathrm{e}^{-i q y}A^{ext}_{0}(y), \quad q=q_x.
\end{equation}
From (\ref{1.4}) and (\ref{1.5}) with account of (\ref{1.2}) for the external Coulomb source  (\ref{0.1}) one obtains the following expression for the induced charge density (here and henceforth $m \to 1$):
\beq
\label{1.6}
\begin{aligned}
\r^{(1)}_{vac}(x)= Z \a |e|\, \frac{2}{\pi^2} \int\limits_{0}^{+\infty}d q\,\cos (q x) &\[1-\frac{2}{q\sqrt{1 +(q/2)^2}}\,\mathrm{arcsinh}\(\frac{q}{2}\) \]\times
\\
&\times \[ \sin(q a)\(\frac{\pi}{2}-\mathrm{Si}(q a) \)-\cos(q a)\mathrm{Ci}(q a)\],
\end{aligned}
\eeq
with $\mathrm{Si}(x)$ and $\mathrm{Ci}(x)$ being the integral sine and cosine functions.
From (\ref{1.3}) via (\ref{1.6}) one finds  the lowest-order vacuum energy
\beq
\label{1.7}
\begin{aligned}
E^{(1)}_{vac}=(Z \a)^2\frac{2}{\pi^2}\int\limits_{0}^{+\infty}d q\, \[ \sin(q a)\(\frac{\pi}{2}-\mathrm{Si}(q a) \)-\cos(q a)\mathrm{Ci}(q a)\]^2\times \\
\times \[1-\frac{2}{q\sqrt{1+(q/2)^2}}\mathrm{arcsinh}\(\frac{q}{2}\) \].
\end{aligned}
\eeq
From (\ref{1.6}) it could be easily seen that within perturbation theory, the total induced vacuum charge vanishes
\beq
\label{1.8}
Q^{(1)}_{vac}=\int \! dx \ \r^{(1)}_{vac}(x)=0 \ .
\eeq
It would be worth-while to note, that although in this case the relation (\ref{1.8}) is an obvious consequence of the explicit form  of perturbative vacuum density (\ref{1.6}), actually the status of the relation (\ref{1.8}) turns out to be quite serious. Namely, it should be considered as a crush-test for the correct calculation of $\r_{vac}(x)$, since  without nontrivial topology or asymptotics of the external field and/or some special boundary conditions  in the subcritical region with $Z<Z_{cr,1}$ the total induced charge should vanish \cite{25,26}. At the same time, for $Z>Z_{cr,1}$, due to nonperturbative effects, caused by discrete levels diving into lower continuum,   the vacuum charge becomes nonzero \cite{9,10,11,13}, and in what follows we will show how the latter circumstance shows up in the behavior of vacuum energy in the overcritical region.

\section{Wichmann-Kroll Contour Integration in 1+1 D}

The most efficient nonperturbative approach to calculation of vacuum density $\r_{vac}(x)$ is based on the  Wichmann-Kroll (WK) method \cite{21,22,23,24} (see also Ref.~\cite{26} and references therein).
The starting point of WK method is the following expression for the induced charge density:
\beq
\label{3.1}
\r_{vac}(x)=-\frac{|e|}{2}\(\sum\limits_{E_{n}<E_{F}} \p_{n}(x)^{\dagger}\p_{n}(x)-\sum\limits_{E_{n}\geqslant E_{F}} \p_{n}(x)^{\dagger}\p_{n}(x) \),
\end{equation}
where  in such problems with external Coulomb source $E_F$ should be chosen at the threshold of the lower continuum, i.e. $E_F=-1$, while $E_{n}$ and $\p_n(x)$ are the eigenvalues and eigenfunctions of corresponding  DC spectral problem.

The essence of WK method is that the vacuum density
(\ref{3.1}) can be calculated by means of  the trace of the Green function for DC spectral problem,  defined as
\beq
\label{3.2}
\(- i \,\a\, \pd_{x}+V(x)+\b -\e \)G(x,x';\e)=\d(x-x'),
\eeq
where it is convenient to take the Dirac matrices as $\a=\s_{y}$, $\b=\s_{z}$.

\begin{figure}[h!]
\centerline{\includegraphics[scale=0.2]{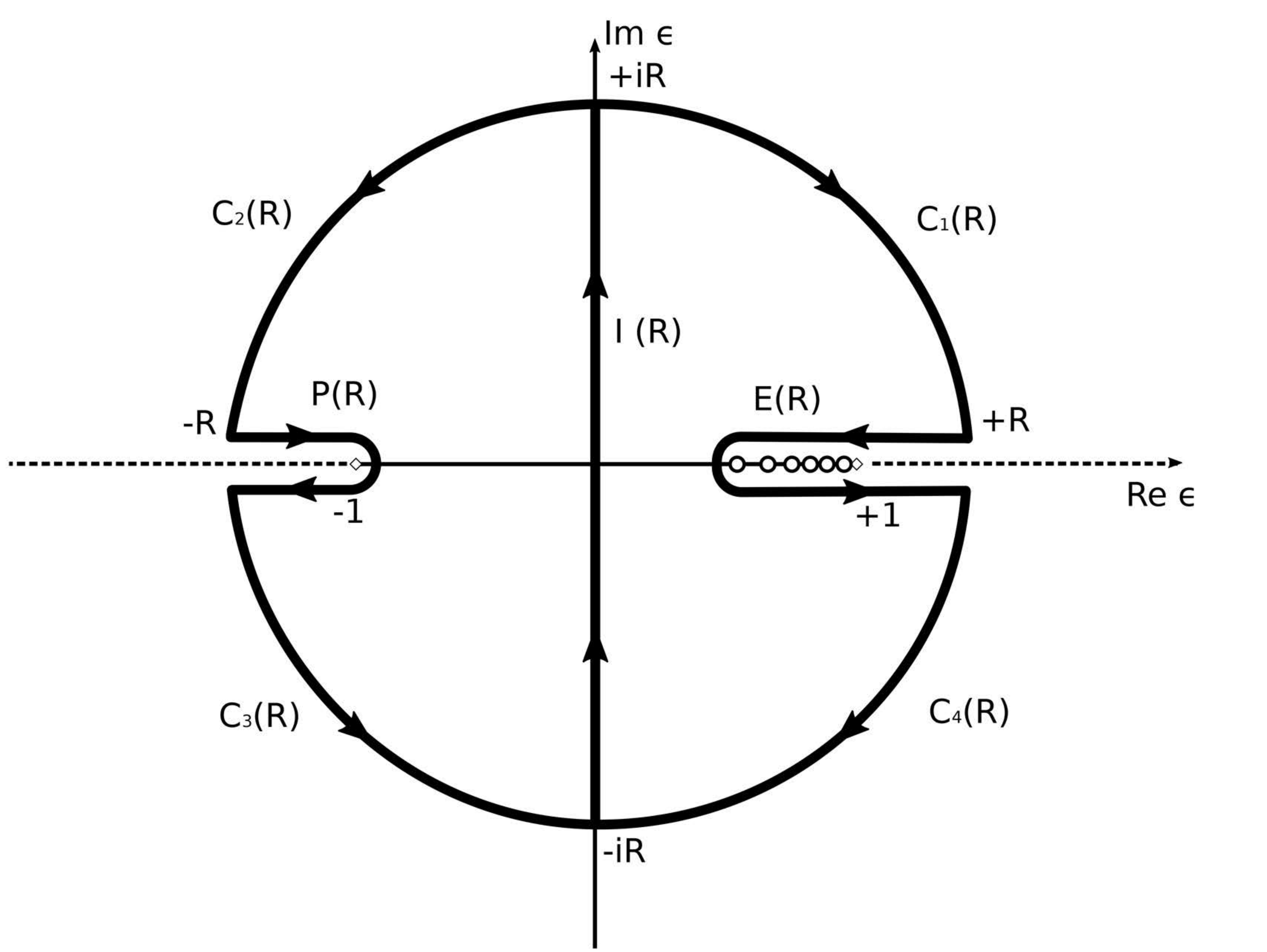}}
\caption{Special contours in the complex energy plane, used for representation of the vacuum charge density via contour integrals. The direction of contour integration is chosen in correspondence with (\ref{3.3}). \label{fig2}}
\end{figure}

The formal solution of (\ref{3.2}) should be written in the form
\beq
\label{3.3}
G(x,x';\e)=\sum\limits_{n}\frac{\p_{n}(x)\p_{n}(x')^{\dagger}}{E_{n}-\e} \ .
\eeq
Following Refs.~\cite{21} and~\cite{22},  the vacuum density  is expressed via integration $\tr G$ along contours $P(R)$ and $E(R)$ on the first sheet of the Riemann energy surface  (Fig.~\ref{fig2})
\beq
\label{3.4}
\r_{vac}(x)=-\frac{|e|}{2} \lim_{R\rightarrow \infty}\( \frac{1}{2\pi i}\int\limits_{P(R)}d\e\, \mathrm{Tr}G(x,x;\e)+\frac{1}{2\pi i}\int\limits_{E(R)}d\e\, \mathrm{Tr}G(x,x;\e) \) \ .
\eeq
Proceeding further, the trace of  Green function is represented as
\beq
\label{3.5}
\mathrm{Tr}G(x,x;\e)=\frac{1}{J(\e)}\p_{L}(x)^{\mathrm{T}} \p_{R}(x),
\eeq
with $\p_{L}(x)$, $\p_{R}(x)$ being the regular at  $\pm \infty$  solutions of DC problem, while $J(\e)$ is their Wronskian
\beq
\label{3.6}
J(\e)= \p_{L,2}(x) \p_{R,1}(x) -\p_{L,1}(x) \p_{R,2}(x).
\eeq
It should be noted that actually $J(\e)$ is nothing else, but the Jost function of  DC problem: the real-valued zeros of $J(\e)$ lie on the first sheet in the interval $-1 \leq \e <1$ and coincide with discrete  levels $E_n$, while the complex ones reside on the second sheet with negative imaginary part of the wavenumber $k=\sqrt{\e^2-1}$, and for $\mathrm{Re}\, k >0$ give rise to elastic resonances.

To construct the Green function,  let us consider first the DC spectral problem, which takes the form
\beq \label{3.61}
\vf'=\left[ \e+1-V(x)\right] \c \ , \quad \c'=-\left[ \e-1-V(x))\right] \vf \ ,
\eeq
with $\vf$ and $\c$ being the upper and lower components of the Dirac wave function. For the potential (\ref{0.1}) the system  (\ref{3.61}) has been considered  in detail in Refs.~\cite{17} and~\cite{27}.
Following Ref.~\cite{27}, for $x>0$  the independent solutions of the Dirac equation are chosen in the form
\beq\label{3.7}
\F(x)=\begin{pmatrix}
\F_1(x;\e) \\
\F_2(x;\e)
\end{pmatrix},\,
\P(x)=\begin{pmatrix}
\P_1(x;\e) \\
\P_2(x;\e)
\end{pmatrix} \ ,
\eeq
where
\beq
\label{3.8}
\begin{aligned}
\F_{1}(x;\e)&=\sqrt{1+\e}\,\mathrm{e}^{-\g z} (2\g z)^{iQ} \(\frac{Q}{\g}\F(b,c,2\g z)+b\,\F(b+1,c,2\g z) \),
\\
\F_{2}(x;\e)&=\sqrt{1-\e}\,\mathrm{e}^{-\g z} (2\g z)^{iQ} \( -\frac{Q}{\g}\F(b,c,2\g z)+b\,\F(b+1,c,2\g z) \),
\\
\P_{1}(x;\e)&=\sqrt{1+\e}\,\mathrm{e}^{-\g z} (2\g z)^{iQ} \(\P(b,c,2\g z)+\frac{Q}{\g}\P(b+1,c,2\g z) \),
\\
\P_{2}(x;\e)&=\sqrt{1-\e}\,\mathrm{e}^{-\g z} (2\g z)^{iQ}  \(-\P(b,c,2\g z)+\frac{Q}{\g}\P(b+1,c,2\g z) \),
\end{aligned}
\end{equation}
with $\F(b,c,x)$, $\P(b,c,x)$ being the confluent hypergeometric functions of  the first and second kinds correspondingly \cite{28},
\beq
\label{3.9}
\begin{aligned}
\g=\sqrt{1-\e^2}, \quad Q=Z \a,  \quad b=iQ - \frac{Q \e}{\g}, \quad c=1+i 2 Q, \quad z=x+a \ .
\end{aligned}
\eeq
$\p_L(x)$ and $\p_R(x)$ should be chosen as such linear combinations of solutions (\ref{3.7}), which are regular at  $-\infty$ and $+\infty$  and connected through  spatial inversion $\p_L(x) = \b \p_R(-x)$ (with the latter being the direct consequence of parity conservation in the initial  problem statement with external potential (\ref{0.1})):
\beq
\begin{aligned}
\label{3.10}
\p_{R}(x)&= \Theta(-x) \b \[A\, \F(-x)+B\, \P(-x)\]+\Theta(x) \P(x),\\
\p_{L}(x)&= \Theta(-x)\(\b\P\)(-x)+\Theta(x)\[A\, \F(x)+B\, \P(x)\] \ .
\end{aligned}
\eeq
From the continuity condition at  $x=0$, one finds for the coefficients $A$ and $B$
\beq
\label{3.11}
A=-2\, \frac{\P_{1}(0;\e)\P_{2}(0;\e)}{\[\P(x),\P(x) \]_{0}},\quad B=\frac{\P_{1}(0;\e)\P_{2}(0;\e)+\P_{2}(0;\e)\P_{1}(0;\e)}{\[\P(x),\P(x) \]_{0}},
\eeq
\begin{equation*}
\Bigl(\[f(x),g(x)\]_{a}=f_{2}(a)g_{1}(a)-f_{1}(a)g_{2}(a)\Bigr).
\end{equation*}
The Wronskian of solutions, which enters the expression for $\tr G$, takes the form
\beq
\label{3.11a}
J(\e)=\[\p_{L}(x),\p_{R}(x)\] = -2\, \P_{1}(0;\e)\,\P_{2}(0;\e) \ .
\eeq
From  (\ref{3.7})-(\ref{3.11a}) and (\ref{3.5}) one finds the following expression for $\tr G$:
\beq
\label{3.12}
\mathrm{Tr}G(x,x;\e)=\frac{1}{\[\F(x),\P(x)\]}\(\F(x)^{\mathrm{T}}\P(x) - \frac{1}{2}\( \frac{\F_{1}(0;\e)}{\P_{1}(0;\e)}+\frac{\F_{2}(0;\e)}{\P_{2}(0;\e)}\)  \P(x)^{\mathrm{T}}\P(x)\) \ ,
\eeq
where by construction $\mathrm{Tr}G(x,x;\e)=\mathrm{Tr}G(-x,-x;\e)$, while $\[\F(x),\P(x)\]$, which enters (\ref{3.12}), equals to
\beq
\label{3.13}
\[\F(x),\P(x) \]= 2 \g\,  \G(c)/\G(b) \ .
\eeq
In the next step one finds the asymptotics of $\tr G$ on the arcs  $C_1(R)$ and $C_2(R)$ in the upper half-plane (Fig.~\ref{fig2})
\beq
\label{3.14}
\tr G (x,x;\e) \to i + {i \over 2 \e^2} - {Q \over |x|+a } {i \over \e^3} + \mathrm{O}(|\e|^{-4}) \  , \quad 0<\mathrm{Arg}\, \e<\pi \ ,  \quad |\e|\to \inf \ ,
\eeq
and on the arcs  $C_3(R)$ and $C_4(R)$ in the lower half-plane
\beq
\label{3.15}
\tr G (x,x;\e) \to -i - {i \over 2 \e^2} + {Q \over |x|+a } {i \over \e^3} + \mathrm{O}(|\e|^{-4}) \  ,  \quad -\pi<\mathrm{Arg}\, \e<0 \ , \quad |\e|\to \inf \ .
\eeq
 There follows from (\ref{3.14}) and  (\ref{3.15}) that the  integration along the  contours $P(R)$ and $E(R)$  in (\ref{3.4}) could be reduced to the imaginary axis (see Fig.~\ref{fig2}), whence one finds the final expression for the vacuum charge density
\beq
\label{3.16}
\r_{vac}(x)=\frac{|e|}{2 \pi i} \int\limits_{-i\infty}^{+i\infty}d\e\, \mathrm{Tr}G(x,x;\e)=\frac{|e|}{2 \pi} \int\limits_{-\infty}^{+\infty}d y\, \mathrm{Tr}G(x,x;iy) \ .
\eeq
In the case when there exist negative discrete levels with $-1\leq E_{n}<0$, instead of (\ref{3.16}) one gets
\beq
\label{3.17}
\r_{vac}(x)=|e|\[\sum\limits_{-1\leq E_{n}<0}\p_{n}(x)^{\dagger}\p_{n}(x)+\frac{1}{2 \pi} \int\limits_{-\infty}^{+\infty}d y\, \mathrm{Tr}G(x,x;iy)\].
\eeq
Proceeding further, let us mention the general properties of $\tr G$ under the change of the sign of external field $(Q \to -Q)$ and complex conjugation
\beq
\label{3.18}
\mathrm{Tr}G_Q(x,x;\e)=-\mathrm{Tr}G_{-Q}(x,x;-\e) \ , \quad \mathrm{Tr}G(x,x;\e)^\ast=\tr G(x,x;\e^\ast) \ ,
\eeq
and their direct consequence
\beq
\label{3.19}
\mathrm{Tr}G_Q (x,x;i y)^{\ast}=-\mathrm{Tr}G_{-Q}(x,x;i y) \ .
\eeq
There follows from  (\ref{3.19}), that $ \mathrm{Re}\[\mathrm{Tr}G_Q (x,x; i y)\]$ is an odd function in $Q$ and an even one in $y$,
whereas $ \mathrm{Im} \[\mathrm{Tr}G_Q (x,x; i y)\]$, on the contrary, is even in $Q$ and odd in $y$. Therefore,  actually $\r_{vac}(x)$ is determined via $ \mathrm{Re} \[\mathrm{Tr}G_Q (x,x; i y)\]$ and so is definitely a real quantity, odd in $Q$. In the purely perturbative region, the representation of $\r_{vac}(x)$ as an odd series in powers of external field follows directly from the Born series  for Green function $G=G^{(0)}+G^{(0)} (-V) G^{(0)} + G^{(0)} (-V) G^{(0)} (-V) G^{(0)} + \dots $, whence
\beq
\label{3.191}
\mathrm{Re}\, \mathrm{Tr}G (x,x; i y)=\sum\limits_{k=0} \mathrm{Re}\, \mathrm{Tr}\[ G^{(0)} \(-V G^{(0)}\)^{2k+1}(x,x; i y) \]  \ ,
\end{equation}
where $G^{(0)}$ is the free Green function. At the same time, in presence of negative discrete levels and, moreover, in the overcritical region with $Z>Z_{cr,1}$, the oddness in $Q$ property of $\r_{vac}$  maintains \cite{22}, but now the dependence on external field cannot be described by a power series (\ref{3.191}) any more, since  there appear in $\r_{vac}$ certain additional, essentially nonperturbative, hence nonanalytic in $Q$ components.

 The expression for the vacuum density, given in (\ref{3.16}) and (\ref{3.17}), does not be automatically consistent with the  requirement of  total induced charge vanishing for $Z<Z_{cr,1}$. Moreover, it could be easily found from the explicit asymptotics of $\tr G$ (\ref{3.14}) and (\ref{3.15}), that $ \mathrm{Re} \[\tr G(x,x; i y)\]$  for $|y| \to \inf$ behaves as $Q/(|x|+a) \times |y|^{-3}$, hence, the integral $\int \! dy \ \mathrm{Re} \[\tr G (x,x; i y)\]$ converges uniformly in $x$. Proceeding further, one finds that, since $ \mathrm{Re} \[\tr G (x,x; i y)\]$  for $|x| \to \inf$ behaves like $Q/(1+y^2)^{3/2} \times |x|^{-1}$, the nonrenormalized   $\r_{vac}(x)$ decreases for $|x| \to \inf$ as $1/|x|$, and so the corresponding induced charge diverges logarithmically.

The general result, obtained in Ref.~\cite{22} via expansion of $\r_{vac}(x)$ in powers of $Q$, and which is valid for any number of spatial dimensions, is that all the divergences of $\r_{vac}$ originate from the lowest-order graph (Fig.~\ref{fig1}) only, while the next-to-leading orders are finite (see also Ref.~\cite{26} and references therein). So the calculation of renormalized vacuum density $\r^{R}_{vac}$ implies, that the terms of order $Q$ should be extracted from the expression for $\tr G$  (\ref{3.12}) and replaced by  $\r^{(1)}_{vac}$ (\ref{1.6}). For these purposes, one finds first the component of the vacuum density $\r^{(3+)}_{vac}$, defined as
\beq \label{3.20}
\r_{vac}^{(3+)}(x)=|e|\[\sum\limits_{-1\leq E_{n}<0}\p_{n}(x)^{\dagger}\p_{n}(x)+\frac{1}{2 \pi} \int\limits_{-\inf}^{+\inf} \! dy\, \(\tr G(x,x;iy)- \tr G^{(1)}(x,x;iy)\)\],
\eeq
where
$G^{(1)}=\left. Q\, \pd G /\pd Q \right|_{Q=0}$ and coincides with the first Born approximation $G^{(0)} (-V) G^{(0)}$. For the external source (\ref{0.1}), the explicit form of $G^{(1)}$ reads
\beq \label{3.21}
\begin{aligned}
\tr G^{(1)}(x,x;i y)&=\frac{Q}{\tg^2}\[\mathrm{e}^{-2\tg(|x|+a)}\(\mathrm{Ei}(2\tg(|x|+a))-\mathrm{Ei}(2\tg a)-\mathrm{e}^{4\tg a} \mathrm{Ei}(-2\tg a) \)-\right.\\
&\left.-\mathrm{e}^{2\tg(|x|+a)}\mathrm{Ei}(-2\tg (|x|+a)) \] \ ,
\end{aligned}
\eeq
with $\tg=\sqrt{1+y^2}$ and $\mathrm{Ei}(x)$ being the integral exponent.

For $Z < Z_{cr,1}$ the integral charge coming from $\r_{vac}^{(3+)}(x)$ vanishes. Without negative discrete levels, which prevent from analytic continuation,  this statement can be proved via transition into the complex $x-$plane, where $\r_{vac}^{(3+)}(x)$, represented through the converging integral  $(|e|/2\pi) \int \! dy \ \(\tr G(x,x;iy)-\tr G^{(1)}(x,x;iy)\)$, turns out to be an analytic function of $x$ with a cut, that appears from the Tricomi function  in $\tr G$, and that could be always directed along the negative imaginary axis. So the integral induced charge $Q_{vac}^{(3+)}=\int \! dx \  \r_{vac}^{(3+)}(x)$ can be expressed via contour integral along the arc of great circle in the upper half-plane. The latter vanishes exactly, what could be easily checked by direct calculation of the asymptotics $\r_{vac}^{(3+)}(x)$. More concretely, the asymptotics of  $\tr G(x,x;i y)$ in the upper half-plane for $\mathrm{Re}\,x>0$ takes the form
\beq \label{3.22}
\tr G(x,x;i y) \to { i y \over (1+y^2)^{1/2} } + {Q \over (1+y^2)^{3/2}} {1 \over x+a } + \mathrm{O}({1 \over x^2}) \ , \quad \mathrm{Re}\,x>0 \ , \quad |x| \to \inf \ .
\eeq
The leading term in the asymptotics (\ref{3.22}) is purely imaginary, even in  $Q$ and odd in $y$, and therefore disappears by integration over $dy$,  the next-to-leading odd in $Q$ term is canceled by $\tr G^{(1)}$, while the remaining terms vanish as $\mathrm{O} (1/x^2)$. For $\mathrm{Re}\,x<0$ the asymptotics $\r_{vac}^{(3+)}(x)$ is found from the reflection symmetry $f^\ast(x)=f(-x^\ast)$, and so on the whole great circle in the upper half-plane $\r_{vac}^{(3+)}(x)$ decreases uniformly as $\mathrm{O} (1/x^2)$.  This result confirms once more the conclusion, that all the divergences in $\r_{vac}(x)$ originate from the terms linear in $Q$.

So the final answer for the renormalized induced charge density reads
\beq\label{3.23}
\r^{R}_{vac}(x)=\r_{vac}^{(1)}(x)+\r_{vac}^{(3+)}(x) \ ,
\eeq
where $\r_{vac}^{(1)}$ is the perturbative renormalized density (\ref{1.6}), calculated via lowest-order diagram (Fig.~\ref{fig1}). Such expression for  $\r^{R}_{vac}$ provides vanishing of the total vacuum charge for $Z<Z_{cr,1}$. In presence of negative discrete levels,  vanishing of the total charge for $Z<Z_{cr,1}$ follows from model-independent arguments,  based on the initial expression for the vacuum density (\ref{3.1}). The latter means that any change  of integral induced charge is possible for $Z>Z_{cr,1}$ only, when certain discrete levels dive into lower continuum, and each diving level yields the change of integral charge by $(-|e|)$.  Another way to achieve the same conclusion follows from the behavior of the integral over the imaginary axis $I(x)=(1/2\pi)\, \int \! dy \ \tr G(x,x;iy)$, which enters the expressions (\ref{3.16}) and (\ref{3.17}) for $\r_{vac}(x)$, under such infinitesimal variation of external source, when the initially positive, infinitely close to zero level $\p_n(x)$ becomes negative. Then the corresponding pole of the Green function undergoes an infinitesimal displacement along the real axis and crosses zero too, what yields the change in $I(x)$ equal to the residue at $\e=0$, namely,  $\D I(x)=\left. -\p_{n}(x)^{\dagger}\p_{n}(x) \right|_{\e=0}$. So the integral induced charge, associated with $I(x)$, loses one unit, whenever there appears the next negative discrete level. Until this negative level exists, this loss of unit charge is compensated by the corresponding term, coming from the sum $\sum \p_{n}(x)^{\dagger}\p_{n}(x)$ over negative discrete levels in (\ref{3.17}). However, as soon as this level dives into lower continuum, the corresponding term in the sum over negative levels disappears, and so the total induced charge loses one unit of $|e|$.

A more detailed picture of resulting changes in $\r^R_{vac}(x)$ is  similar to that considered in Refs.~\cite{9}\nocite{10}--\cite{11}, \cite{13}, \cite{25} for 3+1 D by means of Fano approach to the autoionization in atomic physics \cite{29}. The main result is that whenever the level $\p_n(x)$ dives into lower continuum, the change in the vacuum density takes the form
\beq\label{3.24}
\D\r_{vac}(x)=-|e| \p_n(x)^{\dagger}\p_n(x) \ .
\eeq
It should be noted  that this approach uses some approximations too, and so the expression (\ref{3.24}) turns out to be exact only in the vicinity of corresponding $Z_{cr}$. The correct way of calculation $\r_{vac}^R(x)$  for all regions of $Z$ should be based on relations  (\ref{3.20}) and (\ref{3.23}) with subsequent control of expected  integer value of the induced charge via direct integration of $\r_{vac}^R(x)$.
\begin{figure}[h!]
\centerline{\includegraphics[scale=0.45]{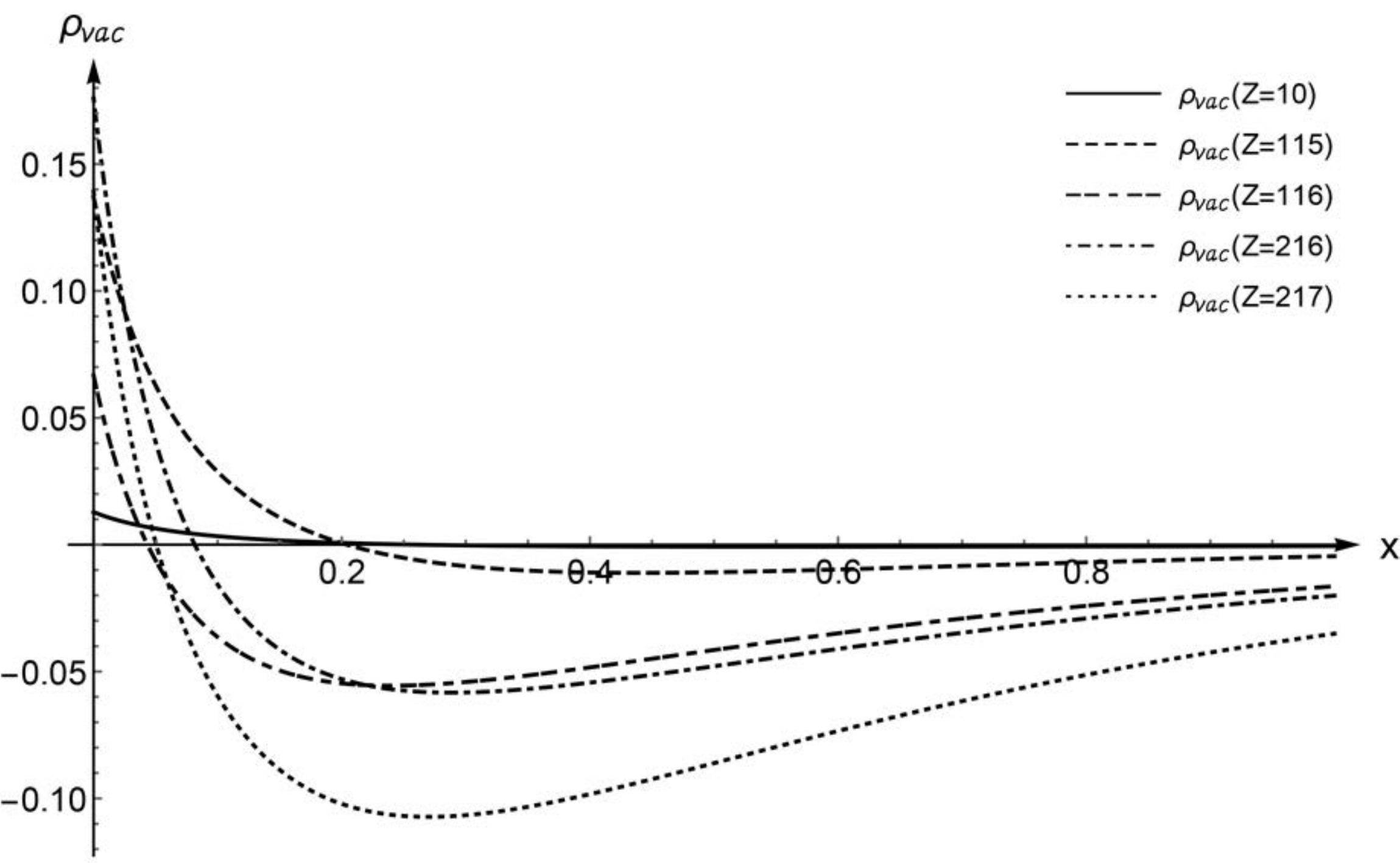}}
\caption{$\r_{vac}^R(x)$ for external potential (\ref{0.1}) for $a=0.1$ and $Z=10, \ 115, \ 116, \ 216, \ 217 $.\label{fig3}}
\end{figure}

\begin{figure}[h!]
\centerline{\includegraphics[scale=0.45]{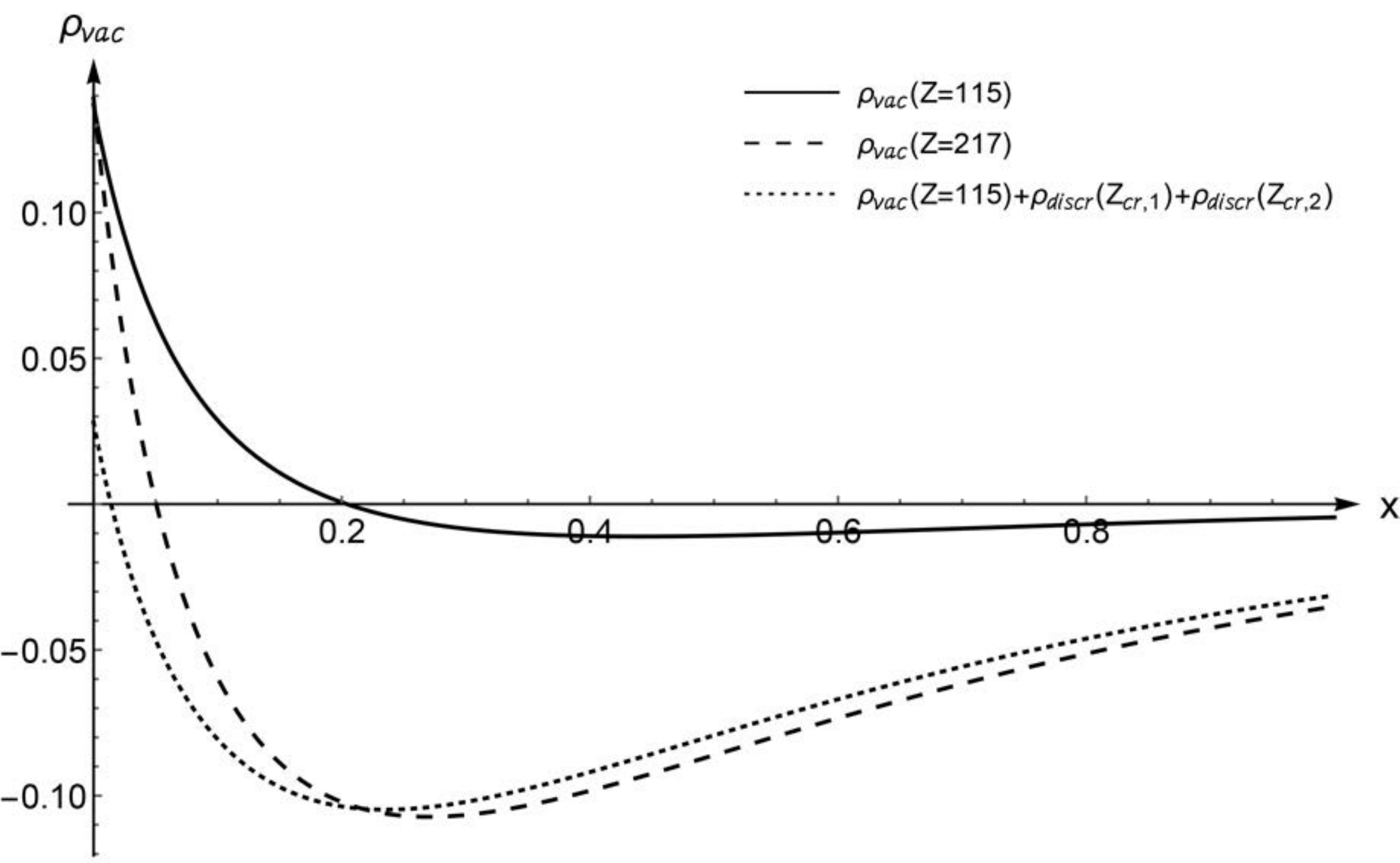}}
\caption{The change of $\r_{vac}^R(x)$ for $a=0.1$ by transition through two fist $Z_{cr}$. The sum of vacuum density for $Z=115$ and of two first discrete levels, taken at the threshold of the lower continuum, does not   reproduce $\r_{vac}^R(x)$ for  $Z=217$ no longer.\label{fig4}}
\end{figure}

An illustration for such a picture is given in Figs.~\ref{fig3}, \ref{fig4} for $a=0.1$ in (\ref{0.1}). Fig.~\ref{fig3} shows the renormalized vacuum density in the purely perturbative regime for $Z=10$, thereafter for  $Z=115$, when the first $Z_{cr,1}\simeq 115.999$ is not  reached yet, for $Z=116$, when the first (even) discrete level has just dived into lower continuum, afterwards for $Z=216$, when the second critical $Z_{cr,2}\simeq 216.258$ is not reached yet, and, finally, for $Z=217$,  i.e. just after diving of the second (odd) discrete level into lower continuum. Herewith the direct numerical integration confirms that the total vacuum charge for $Z=10, \ 115$ equals to zero, for $Z=116, \ 216$ equals to $(-|e|)$, while for $Z=217$ equals to $(-2 |e|)$, correspondingly. The critical charges are found from \cite{27}
 \beq \label{3.251}
K_{1+2 i Q}\( \sqrt{8 Q a}\) + K_{1-2 i Q}\( \sqrt{8 Q a}\)=0
\eeq
for even levels and
 \beq \label{3.252}
K_{2 i Q}\( \sqrt{8 Q a}\) =0
\eeq
for odd.

Fig.~\ref{fig4} demonstrates that in the overcritical region with $Z>Z_{cr,1}$  the changes of $\r_{vac}^R(x)$ for increasing $Z$  proceed not only in a step-like manner due to  vacuum shell formations, originating from  levels diving into lower continuum, but  via permanent deformations in the density of states in both continua and evolution of discrete levels too. Namely, it displays the total changes in $\r_{vac}^R(x)$ by transition through two first  $Z_{cr}$. The sum of vacuum density for $Z=115$ and of two first discrete levels, taken at the threshold of the lower continuum, does not  reproduce $\r_{vac}^R(x)$ for  $Z=217$ no longer.

\section{Nonperturbative Effects in Vacuum Energy for $Z > Z_{cr,1}$}

Nonperturbative polarization effects, manifesting in $\r_{vac}^R(x)$ for $Z>Z_{cr,1}$ through formation of localized vacuum shells, give rise to corresponding nonperturbative changes in $E_{vac}$.  As the formation of shells itself, this effect turns out to be essentially nonperturbative, but hitherto has not been considered in detail, since it was assumed that in the overcritical region the main contribution to  $E_{vac}$ should be produced by perturbative effects (see e.g. Ref.~\cite{11} and references therein). It should be specially marked that this effect cannot be evaluated  directly via $\r_{vac}(x)$, since in the nonperturbative region there do not work neither  (\ref{3.191}), nor  perturbative methods of $E_{vac}$ restoration from $\r_{vac}(x)$.  Let us also mention that the growth rate of the shells total number and the increase of the shell effect in $E_{vac}$ with increasing $Z$ depend very strongly on the number of spatial dimensions. In 1+1 D this rate is minimal, that is why the behavior of $E_{vac}$ in the overcritical region depends not as much on shells, but on the  renormalization term combined  with  degradation of the perturbative component in the nonrenormalized $E_{vac}$ with increasing $Z$, what itself turns out to be an essentially nonperturbative effect.

The nonperturbative approach to vacuum energy calculation starts from the expression \cite{9,10,11,25}
\beq
\label{4.1}
E_{vac}=\frac{1}{2} \(\sum\limits_{E_{n}<E_{F}} E_{n}-\sum\limits_{E_{n} \geqslant E_{F}} E_{n} \),
\eeq
which follows from the Dirac Hamiltonian, written in the invariant under charge conjugation form, and is defined up to the choice of the energy origin. In (\ref{4.1}), even in absence of the external field, vacuum energy is negative and divergent. At the same time, the starting expression for the vacuum density (\ref{3.1}) vanishes identically for $A^{ext} = 0$. From this point of view, the most natural way is to normalize $E_{vac}$ on the free case. Moreover, in the external fields like (\ref{0.1}) there exists a (infinite) number of bound states. Therefore, in order to keep in $E_{vac}$ the interaction effects only, the electron rest mass should be subtracted from  the energy of each bound state. Thus, in  physically motivated  form the initial expression for $E_{vac}$ should be written as
\beq
\label{4.2}
E_{vac}=\frac{1}{2} \(\sum\limits_{E_{n}<E_{F}} E_{n}-\sum\limits_{E_{n}\geqslant E_{F}} E_{n} \)_{A}-\frac{1}{2} \(\sum\limits_{E_{n}<0} E_{n}-\sum\limits_{E_{n}>0} E_{n} \)_{0}-\frac{1}{2} \sum\limits_{-1\leq E_{n}<1} (-1).
\eeq
$E_{vac}$, defined in such a way, vanishes in absence of the external field, and so is in complete correspondence with the vacuum charge density $\r_{vac}$ (\ref{3.1}).

Variation of  $E_{vac}$ with respect to $A^{ext}_{0}$ leads to  well-known Schwinger result in the form of  vacuum charge density $\r_{vac}(x)$, combined with  additional term, caused by the nonperturbative vacuum reconstruction for $Z>Z_{cr,1}$:
\beq\label{4.3}
\d E_{vac}=\int\limits_{-\infty}^{+\infty} d x\,\r_{vac}(x)\,\d A^{ext}_{0}(x)+\d E_{N},
\end{equation}
where $E_{N}=-N$ with $N$ being the number of bound states, dived into lower continuum. $E_{N}$ yields a negative contribution to $E_{vac}$ and has the form of a step-like function \cite{11}. Whenever a discrete level reaches the lower continuum, one unit of the electron rest mass is lost by the vacuum energy via $E_N$. Such  fixed negative jumps in the vacuum energy are treated as indication on phase transition from the neutral vacuum into the charged one, which turns out to be the ground state of the electron-positron field in supercritical Coulomb fields \cite{9,10,11,13,25}.

In the next step  (\ref{4.2}) should be divided into separate contributions from discrete and continuous spectra, applying to the difference of integrals over the continuous spectrum $\(\int \! dk \ \sqrt{k^2+1}\)_A - \(\int \! dk \ \sqrt{k^2+1}\)_0$ the well-known tool, which represents this difference in the form of an integral from the elastic scattering phase $\d(k)$. Such technique has been used quite effectively in calculation of one-loop quantum corrections to the soliton mass in essentially nonlinear field-theoretic models in 1+1 D (see Refs.~\cite{30} and~\cite{31} and references therein). Upon dropping certain almost obvious intermediate steps, the final answer reads
\beq
\label{4.4}
E_{vac}=\frac{1}{2 \pi}\int\limits_{0}^{\infty}\frac{k}{\sqrt{k^2+1}}\d_{tot}(k)\,d k+\frac{1}{2}\sum\limits_{-1\leq E_n<1}\(1-E_{n}\),
\eeq
where $\d_{tot}(k)$ is the sum of  phase shifts for the given wavenumber  $k$ from electron and positron scattering states of both parities.

Such approach to calculation of $E_{vac}$ turns out to be quite effective, since $\d_{tot}(k)$ behaves much better, than each of elastic phases separately, both in IR and UV limits, and  turns out to be automatically an even function of the external field. Moreover, in 1+1 D, for the Coulomb potentials like (\ref{0.1}), the vacuum energy, taken in the form (\ref{4.4}), turns out to be finite without any special UV-renormalization. As it will be shown by direct calculation below,  $\d_{tot}(k)$  is finite for $k \to 0$ and  behaves like $\mathrm{O}(1/k^3)$ for $k \to \inf$, hence, the phase integral in (\ref{4.4}) is always convergent. In turn, the total  bound energy of discrete levels is also finite, since $1-E_n$ behave like $\mathrm{O}(1/n^2)$ for $n \to \inf$.
However, the convergence of $E_{vac}$ after subtraction (\ref{4.2})   does not mean any hidden UV-renormalization, rather it is caused exclusively by  specifics of 1+1 D. Although the subtraction (\ref{4.2}) should be implemented under suitable intermediate UV-regularization,  actually it is nothing else, but the choice of pertinent reference frame for $E_{vac}$.  The need in renormalization of $E_{vac}$ via the lowest-order diagram (Fig.~\ref{fig1}) follows from  the analysis of $\r_{vac}(x)$, performed in the preceding section. The latter shows that without such genuine UV-renormalization  the  induced charge does not  acquire the value that should be expected  from general grounds \cite{25,26}. Another requirement is that for $Z \to 0$ the answer for $E_{vac}$ should coincide with the perturbative result $E_{vac}^{(1)}$, found  from (\ref{1.3})-(\ref{1.7}). Since for $Z \to 0$ the connection between $E_{vac}$ and $\r_{vac}(x)$ is described by  perturbative relations (\ref{1.3})-(\ref{1.5}), it is easy to verify that nonrenormalized $E_{vac}$ does not satisfy this condition. With more details this question is explored below in terms of the renormalization coefficient $\l$ (\ref{4.6}).

Thus, in the way quite similar to $\r_{vac}(x)$, we should pass from $E_{vac}$ to the renormalized vacuum energy $E^{R}_{vac}$. In the form, well-adapted for practical use, $E^{R}_{vac}$ could be represented as
\beq
\label{4.5}
E^{R}_{vac}(Z)=E_{vac}(Z) + \l Z^2 ,
\eeq
where
\beq
\label{4.6}
\l=\lim_{Z_0 \to 0} \frac{E^{(1)}_{vac}(Z_{0})-E_{vac}(Z_0)}{Z_0^2} \ ,
\eeq
while the renormalization coefficient $\l$ depends solely on the profile of the external Coulomb field and, depending on the parameters of the source, could be of arbitrary sign, as well as negligibly small (see below).

Now let us consider the calculation of  $E^{R}_{vac}(Z)$ for the Coulomb source of the form (\ref{0.1}).
For these purposes, the spinor components of solutions of the Dirac equation in the upper and lower ($\pm$) continua for $x>0$ should be chosen in the following form
\beq
\label{4.7}
\begin{aligned}
\F^{\pm}_{1}(x;\e)&=\left\lbrace\begin{matrix}
\sqrt{\e+1} \\
\sqrt{|\e|-1}
\end{matrix}\right\rbrace\, \mathrm{Re}\[\mathrm{e}^{i \x^{\pm}}\,\mathrm{e}^{ikz}(-2ikz)^{i Q} \(i\, \frac{Q}{k} \F_{z}+b\, \F_{z}(b+) \) \], \\
\F^{\pm}_{2}(x;\e)&=\left\lbrace\begin{matrix}
-\sqrt{\e-1}\\
\sqrt{|\e|+1}
\end{matrix}\right\rbrace\, \mathrm{Re}\[i\,\mathrm{e}^{i \x^{\pm}}\,\mathrm{e}^{ikz}(-2ikz)^{i Q} \(-i\,\frac{Q}{k} \F_{z}+b\, \F_{z}(b+) \) \],
\end{aligned}
\eeq
where
\beq\label{4.8}
k=\sqrt{\e^2-1} \ , \quad b=i Q \(1-\e /k \) \ , \quad c=1+2 i Q \ ,\quad z=x+a \ , \nonumber \eeq
\beq\label{4.8a} \F_{z}=\F(b,c,-i2kz) \ , \quad \F_{z}(b+)=\F(b+1,c,-i2kz) \ .
\eeq
The relations for coefficients $\x^{\pm}$ are derived from the conditions of even    $(\F_2(0;\e)=0)$  or odd $(\F_1(0;\e)=0)$ continuation of solutions (\ref{4.7}) to negative half-axis $x<0$. As a result, for  even solutions
\beq\label{4.9}
\mathrm{e}^{i 2\x^{\pm}_{even}}=\mathrm{e}^{-i2k{a}}(2ka)^{-i2Q}\, {\(b\,\F_{a}(b+)\)^{*} + i\, (Q/k) \,\F_{a}^{*} \over b\, \F_{a}(b +)-i\, (Q/k)\, \F_{a} } \ ,
\end{equation}
while for odd
\beq\label{4.10}
\mathrm{e}^{i 2\x^{\pm}_{odd}}=- \mathrm{e}^{-i2k{a}}(2ka)^{-i2Q}\, {\(b\,\F_{a}(b+)\)^{*} - i\, (Q/k) \,\F_{a}^{*} \over b\, \F_{a}(b +) + i\, (Q/k)\, \F_{a} } \ .
\eeq
The solutions of Dirac equation for the discrete spectrum ($-1\leq \e<1$) for $x>0$ should be written as
\beq\label{4.11}
\begin{aligned}
\F_1(x;\e)&=\sqrt{1+\e}\, \mathrm{e}^{-\g z} \mathrm{Re} \[ \mathrm{e}^{i \x}(2\g z)^{i Q} \(\frac{Q}{\g} \F(b,c,2\g z) + b\, \F(b+1,c,2\g z)\) \], \\
\F_2(x;\e)&=\sqrt{1-\e}\, \mathrm{e}^{-\g z} \mathrm{Re} \[\mathrm{e}^{i \x}(2\g z)^{i Q} \(-\frac{Q}{\g}\F(b,c,2\g z) + b\, \F(b+1,c,2\g z)\) \] \ ,
\end{aligned}
\eeq
where $\g$ and $b$ are defined as in (\ref{3.9}). Discrete levels are found from the condition for $x \to \inf$ combined with $\F_2(0;\e)=0$ for even levels and $\F_1(0;\e)=0$ for odd, what gives the equation
\beq \label{4.12}
\mathrm{Im} \[(2 \g a)^{iQ} \(-\frac{Q}{\g}\F(b,c,2\g {a})+b\, \F(b+1,c,2\g {a}) \) \G (b) \G (c^*) \]=0 \ ,
\eeq
for even levels, and
\beq \label{4.13}
\mathrm{Im} \[(2 \g a)^{iQ} \(\frac{Q}{\g}\F(b,c,2\g {a})+b\,\F(b+1,c,2\g {a}) \) \G (b) \G (c^*) \]=0 \eeq
for odd.

\begin{figure}[h!]
\centerline{\includegraphics[scale=0.35]{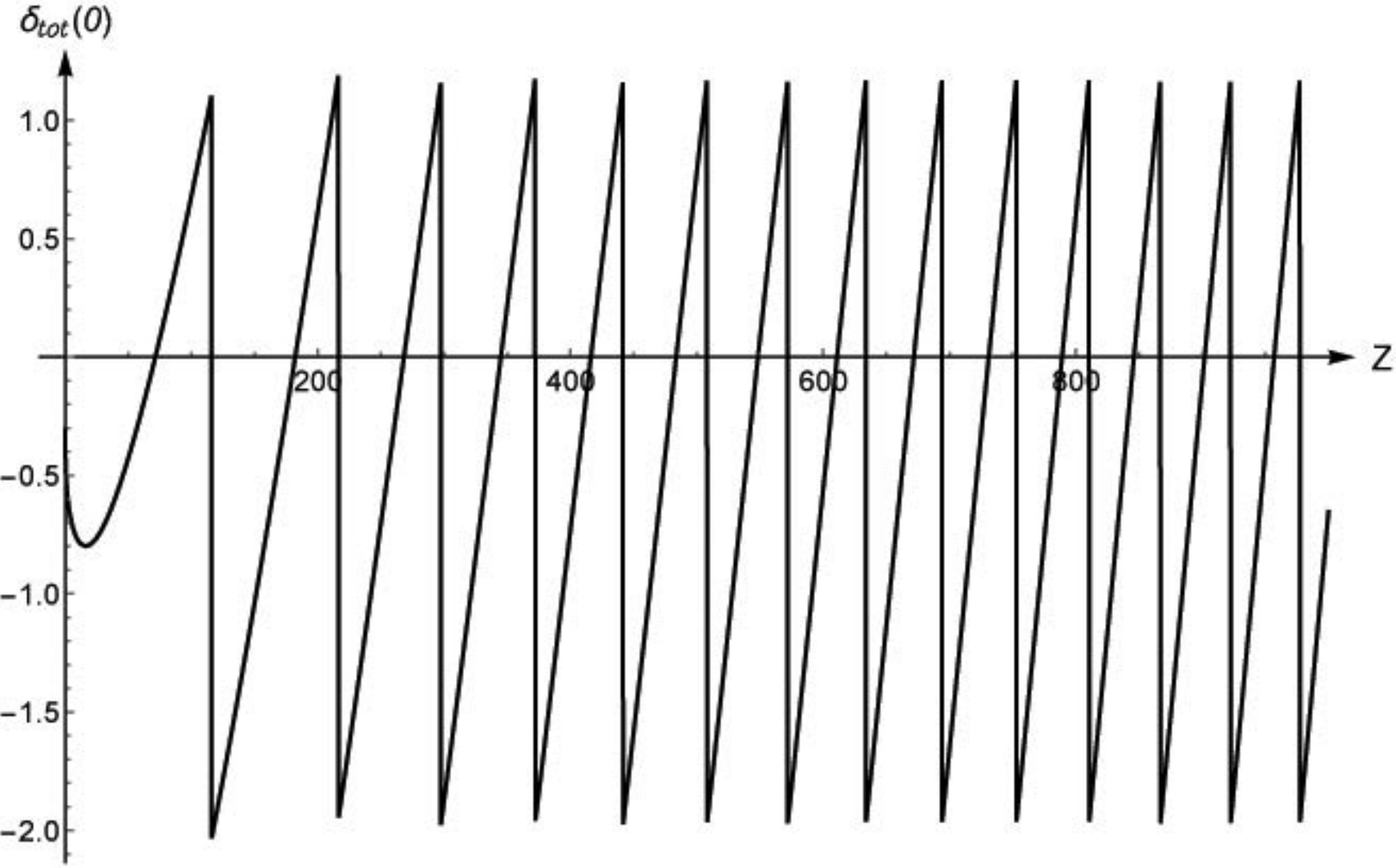}}
\caption{$\d_{tot}(0)$ for $a=0.1$.\label{fig5}}
\end{figure}

Separate phase shifts are found from the asymptotics of solutions (\ref{4.7}) for $x\to +\infty$ and contain Coulomb logarithms $Q\, (\e/k)\, \ln (2 k (|x|+a))$, which in the total phase
\beq \label{4.14}
\d_{tot}(k)=\(\d^{+}_{even}+\d^{+}_{odd}+\d^{-}_{even}+\d^{-}_{odd}\)(k)
\eeq
cancel each other, so in (\ref{4.14}) for separate phases only the regular at  $x \to \inf$ part should be kept, namely
\beq\label{4.15}
\d^{\pm}_\s(k)=\mathrm{Arg}\[\mathrm{e}^{ika + i\f^\pm_\s} \(\mathrm{e}^{\pi Q} \frac{\G(c) \mathrm{e}^{i \l^{\pm}_\s}}{(\e + k)\G[iQ(1+\e/k)]} + \frac{\G(c^*) \mathrm{e}^{-i \l^{\pm}_\s}}{\G[-iQ(1-\e/k)]} \) \] \ ,
\eeq
where $\s= \mathrm{even,\,odd}$, $\f^+_{even}=0$, $\f^-_{even}=\pi$ and $\f^\pm_{odd}=\pm \pi/2$. The  phases (\ref{4.15})  contain for $k \to \inf$ the logarithmic terms $\mp Q\, (\e/k)\, \[\ln (2 k a) -1 \]$, which cancel each other again, and so there remains in the asymptotics of $\d_{tot}(k)$  only the regular part, decreasing $\sim 1/k^3$. More concretely,  the UV-asymptotics of $\d_{tot}(k)$  takes the form
\beq \label{4.16}
\d_{tot}(k\to \inf) = - { 2 Q^2 \over k^3 a } \, \[1+ {1 + 4 Q^2 \over 12 (k a )^2} \] + \mathrm{O}(1/k^7) \ .
\eeq
It should be mentioned that the calculation of (\ref{4.16}) for reasonable time could be performed by means of symbolic computer algebra only.
 $\d_{tot}(k)$ is finite for $k \to 0$ as well, since all the singular terms in the IR-asymptotics $\pm \[ (1- \ln [Q/k]) Q/k- \pi/4 - k Q  \ln [Q/k]/2 \]$ for $\s=$ even and $\pm \[ (1- \ln [Q/k]) Q/k + \pi/4 - k Q \ln [Q/k]/2 \]$ for $\s=$ odd in the upper and lower continua phases cancel as well.
The exact IR-asymptotics for $\d_{tot}(k)$ is found by means of Taylor representation for confluent hypergeometric functions \cite{28} and reads
\beq \label{4.17}
\begin{aligned}
\d_{tot}(k\to 0) = \mathrm{Arg} \[\( \mathrm{e}^{i \vf^+_{even}} - \mathrm{e}^{- 2 \pi Q} \mathrm{e}^{- i \vf^+_{even}}\) \( \mathrm{e}^{i \vf^+_{odd}} - \mathrm{e}^{- 2 \pi Q} \mathrm{e}^{- i \vf^+_{odd}}\) \right.\times \\
\left.\times \sin \(\vf^-_{even}\) \sin \(\vf^-_{odd}\)\] + \mathrm{O}(k) \ ,
\end{aligned}
\eeq
where
\beq\label{4.18}
\begin{aligned}
&\vf^+_{even}=-\mathrm{Arg} \[- J_{2 i Q}\(\sqrt{ 8 Q a}\) \] \ ,
\\
&\vf^+_{odd}=-\mathrm{Arg} \[ \sqrt{ 2 Q a}\, J_{1+2 i Q}\(\sqrt{ 8 Q a}\)- i Q J_{2 i Q}\(\sqrt{ 8 Q a}\) \] \ ,
\\
&\vf^-_{odd}=-\mathrm{Arg} \[ J_{2 i Q}\(\sqrt{ - 8 Q a}\) \] \ ,
\\
&\vf^-_{even}=-\mathrm{Arg} \[ \sqrt{- 2 Q a} \, J_{1+2 i Q}\(\sqrt{- 8 Q a}\)- i Q J_{2 i Q}\(\sqrt{- 8 Q a}\) \] \ .
\end{aligned}
\eeq
In the considered range of external source parameters, the value of $\d_{tot}(0)$ is uniquely determined through expressions (\ref{4.17})--(\ref{4.18}) and lies always in the interval $-\pi<\d_{tot}(0)<\pi$, what could be checked numerically via explicit restoration of $\d_{tot}(k)$ on the whole half-line $0 \leq k \leq \inf$, starting from UV-asymptotics (\ref{4.16}). The behavior of $\d_{tot}(0)$ as a function of $Z$ is shown in Fig.~\ref{fig5}.

The typical behavior $\delta_{tot}(k)$ as a function of wavenumber $k$ is shown in Fig.~\ref{fig6} for $a=0.1$. For such a size of external source the critical charge for the first even level amounts to $Z_{cr,1}\simeq 115.999$, therefore for  $Z=115$ (Fig.~\ref{fig6}a) the overcritical region is not reached yet, and so $\d_{tot}(k)$ reveals a smooth behavior on the whole half-line $0 \leq k \leq \inf$. For $Z=117$ (Fig.~\ref{fig6}b) the first even level has already dived into lower continuum, thence, there appears in the phase the first and yet sufficiently narrow elastic (positron)  resonance. The origin and features of such resonances in the overcritical region are caused by evolution of the Green function poles, corresponding to discrete levels, into second sheet of the Riemann energy surface after achieving the threshold of lower continuum (the beginning of left cut on Fig.~\ref{fig2}), and have been discussed in detail in Refs.~\cite{9}--\cite{11}, \cite{13}, \cite{25}.  To demonstrate the displacement and broadening of resonances with increasing  $Z$, Fig.~\ref{fig6}c shows  $\d_{tot}(k)$ for $Z=298$: two first resonances, which appear after achieving the lower continuum by the first even ($Z_{cr,1}\simeq 115.999$) and first odd ($Z_{cr,2}\simeq 216.258$) discrete levels, become already sufficiently less pronounced, while the second even level ($Z_{cr,3} \simeq 297.24$) has just dived into the lower continuum, the corresponding pole lies very close to the beginning of the left cut and shows up in the phase as an extremely narrow  low-energy resonance.
\begin{figure}[h!]
\begin{minipage}{0.48\linewidth}
\center{ \includegraphics[scale=0.4]{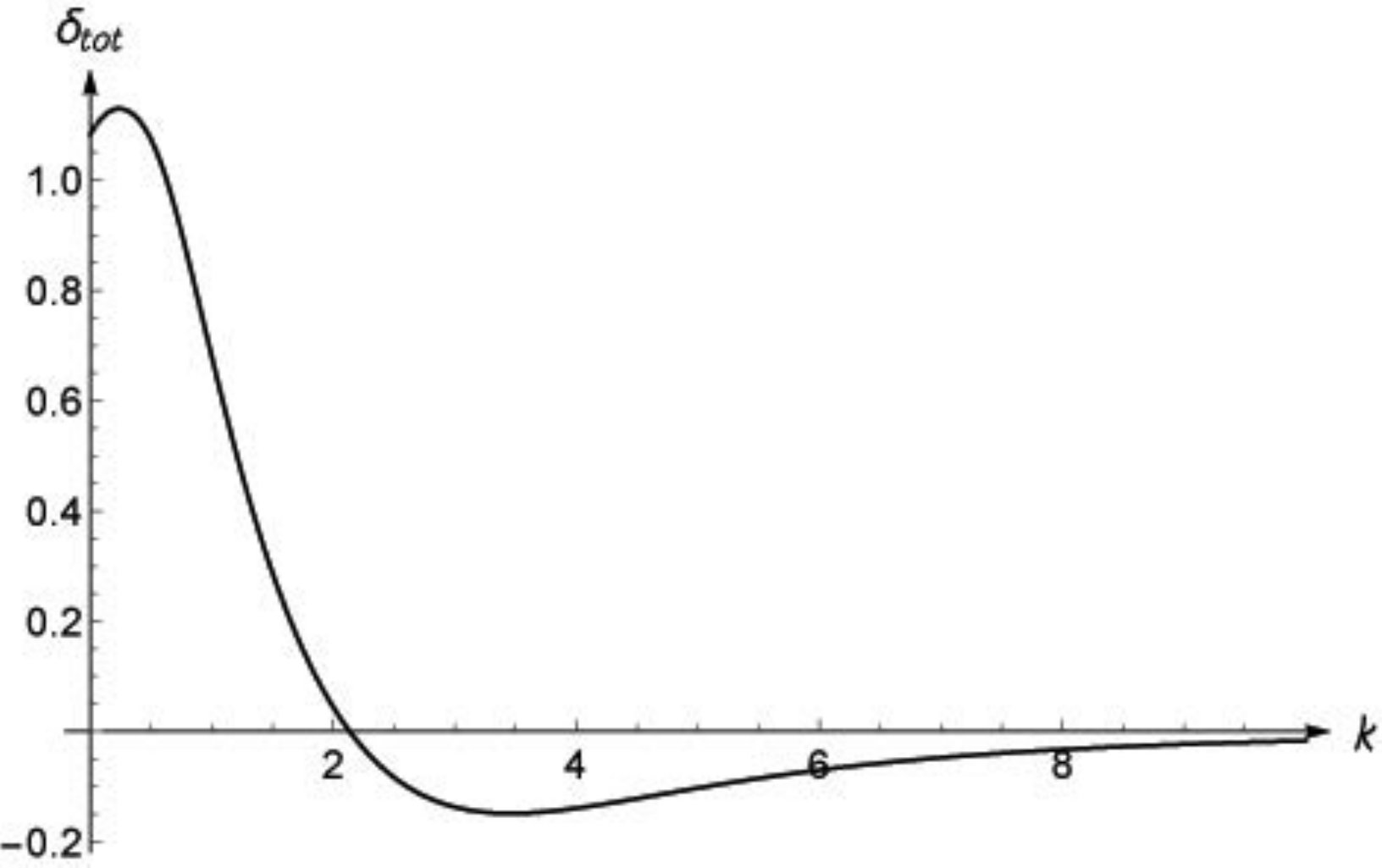} \\ a) }  \\
\end{minipage}
\hfill
\begin{minipage}{0.48\linewidth}
 \center{\includegraphics[scale=0.31]{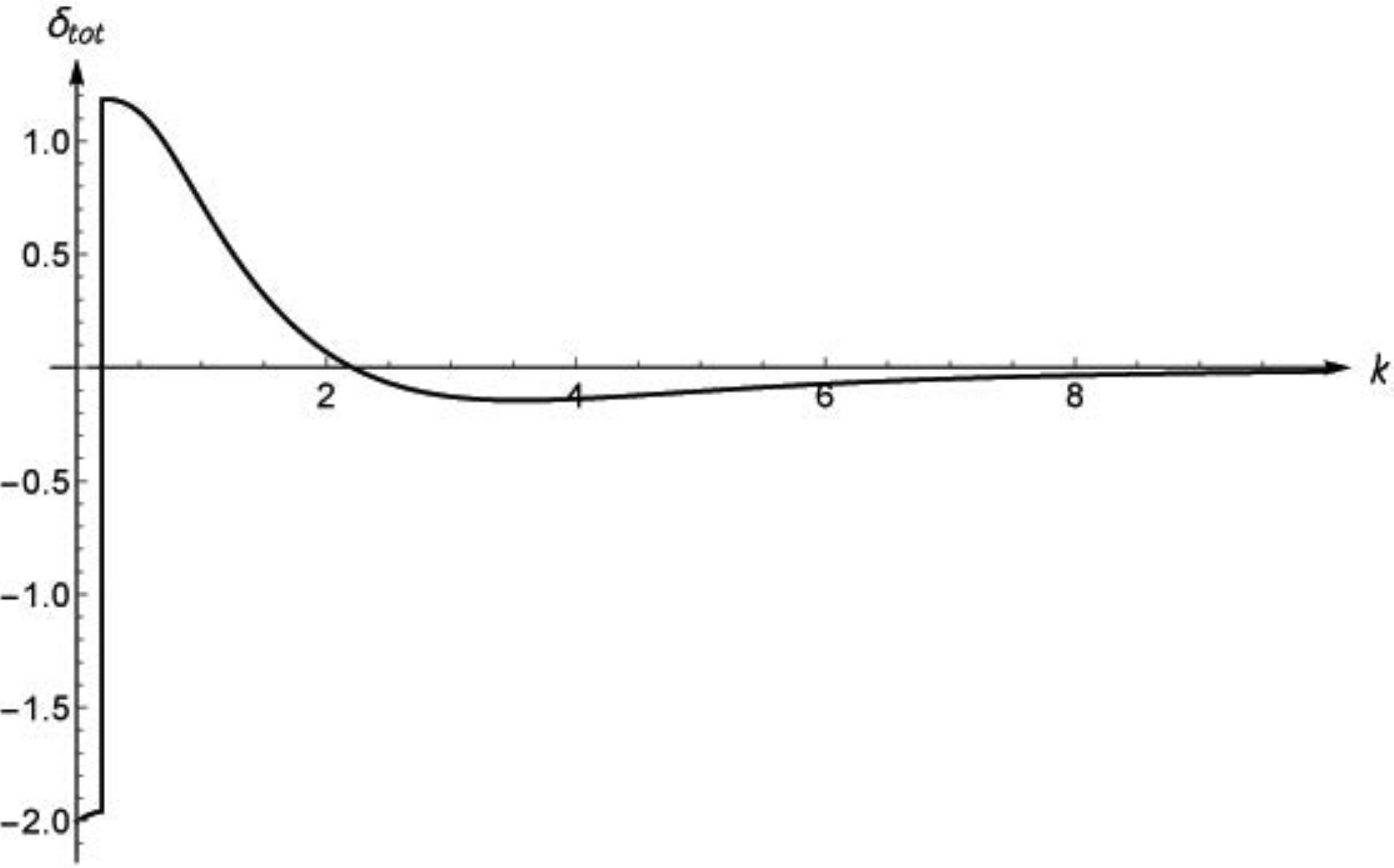} \\ b) }   \\
  \end{minipage}
\end{figure}
\begin{figure}[h!]
\center{\includegraphics[scale=0.35]{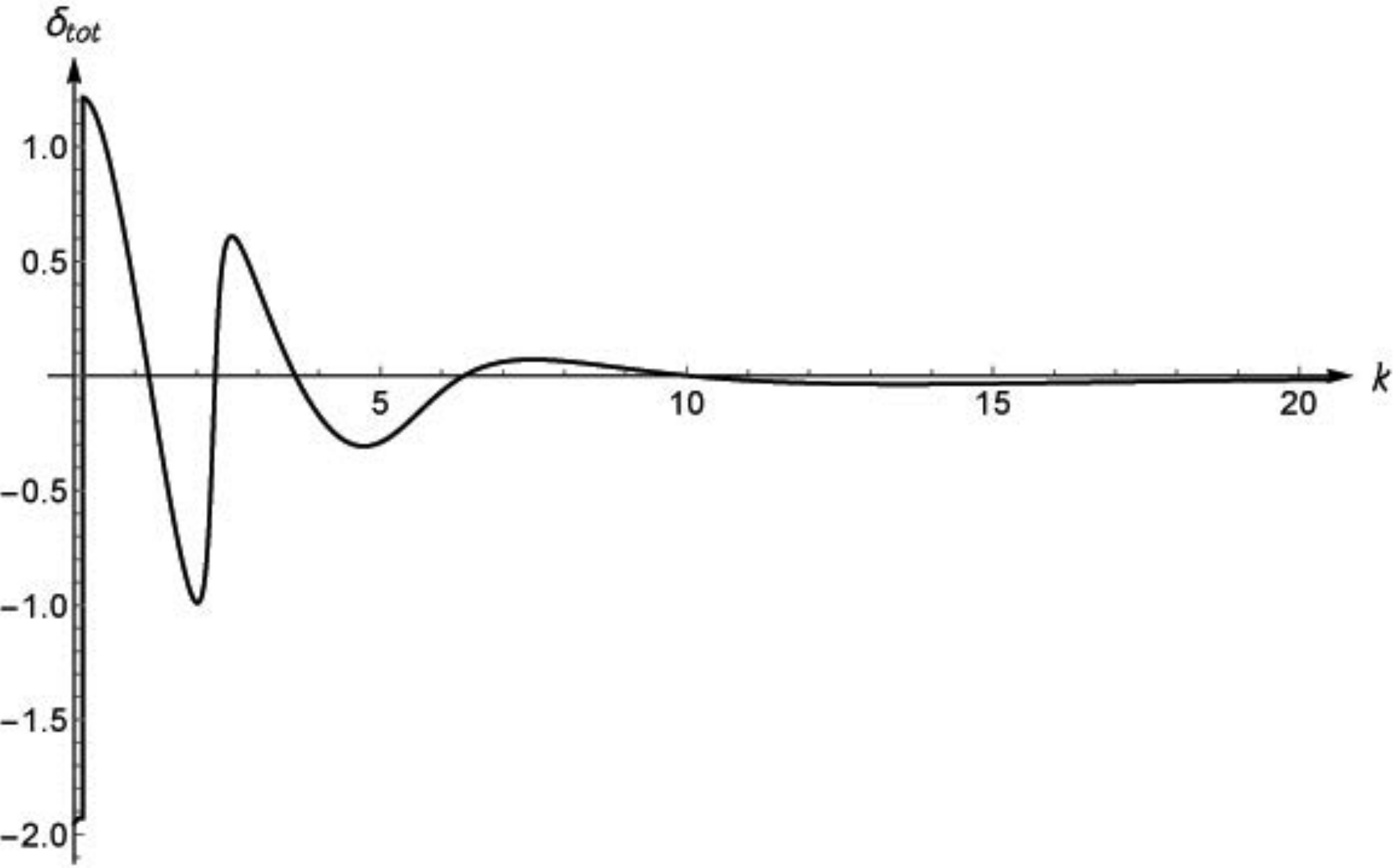} \\ c)}
\caption{$\d_{tot}(k)$ for (a) $Z=115$, \ (b) $Z=117$, \ (c) $Z=298$ and $a=0.1$.\label{fig6}}
\end{figure}\\
As a result,   in (\ref{4.4}) both the phase integral and the total bound energy of discrete levels converge without any additional regularization of Coulomb asymptotics of external potential for $|x|\to \infty$, and so can be evaluated by means of standard numerical recipes. It should be mentioned also that actually the UV-asymptotics (\ref{4.16}) of $\d_{tot}(k)$  is valid for $k a \gg 1$. This circumstance plays a significant role in calculation of the phase integral in (\ref{4.4}), since it allows for a substantial  simplification of integration for asymptotically large $k$.

Now let us specify the main nonperturbative effects in the behavior of phase integral and  total bound energy $\sum_n (1-E_n)$ in (\ref{4.4}) for the overcritical region. As it follows from Fig.~\ref{fig5}, $\delta_{tot}(0)$ oscillates in limits  $-\pi <\d_{tot}(0)<\pi$,  and so each subsequent resonance gives rise to negative jump of the phase equal to $\pi$, thereafter $\delta_{tot}(0)$ increases smoothly  by $\pi$ again. As a result, after emergence of the next resonance,  phase integral acquires a certain negative jump of the derivative, and so becomes an oscillating and smoothly decreasing function of $Z$. On the contrary, discrete levels bound energy grows continuously  with $Z$, since there grows with $Z$  the bound energy of each level, except  $Z_{cr}$, when the corresponding level dives into lower continuum and there appears a jump of  bound energy equal to $(-2 \times mc^2)$. Moreover, in 1+1 D it is indeed the total bound energy, that  dominates in $E_{vac}$ in the overcritical region before renormalization, rather than the phase integral, which gives in $E_{vac}$ a significantly  less contribution. Such behavior of the phase integral and  $E_{vac}$ is a peculiar feature of 1+1 D.

The next factor, affecting the behavior of $E_{vac}^R$ in $O(Z^2)$, is the  renormalization term $\l Z^2$ in (\ref{4.5}). Moreover, in 1+1 D this contribution in $E_{vac}^R$ turns out to be the dominant one in the overcritical region, since the sum of discrete levels and so the non-renormalized $E_{vac}$ grows in this region $\sim Z^p \ , \ 1< p <2$. At the same time, the number of dived levels, hence the number of vacuum shells, increase  $\sim Z^s \ , \ 1< s <2$,  but the correlation between  $p$ and $s$ is not simple, since the relation between  $\r_{vac}$ and $E_{vac}$ in the overcritical region is essentially nonlinear.

For the external source like (\ref{0.1}), there follows directly from the definition of  $\l$ (\ref{4.6}) and dimensional arguments that the asymptotics of $\l(a)$ for both $a \to 0$ and $a \to \inf$ should be  $\sim 1/a$.  For more details, let us represent $\l$ as
\beq
\label{4.19}
\l/\a^2=\l_1-\l_2 ,
\eeq
where $\l_1$ originates from perturbative polarization energy $E^{(1)}_{vac}$, found in (\ref{1.7}),
\beq
\label{4.20}
\l_1= \frac{2}{\pi^2}\int\limits_{0}^{+\infty} \! d q\, \[ \sin(q a)\(\frac{\pi}{2}-\mathrm{Si}(q a) \)-\cos(q a)\mathrm{Ci}(q a)\]^2\[1-\frac{4}{q\sqrt{q^2+4}}\mathrm{arcsinh}\(\frac{q}{2}\) \],
\eeq
while $\l_2$ comes from the  Born term for $\r_{vac}$  (\ref{3.191}), (\ref{3.21})
\beq\label{4.21}
\begin{aligned}
&\l_2= \frac{1}{\pi}\int\limits_{0}^{+\infty} {dy \over 1+y^2} \ \[e^{2 a \sqrt{1+y^2}} \mathrm{Ei}\[-2 a \sqrt{1+y^2}\] \( e^{-2 a \sqrt{1+y^2}} \mathrm{Ei}\[2 a \sqrt{1+y^2}\] + \right.\right.\\
&\left.\left.+ e^{2 a \sqrt{1+y^2}} \mathrm{Ei}\[-2 a \sqrt{1+y^2}\]\) + \int\limits_{0}^{+\infty} {dx \over x+a} \ \( e^{-2 (x+a) \sqrt{1+y^2}} \mathrm{Ei}\[2 (x+a) \sqrt{1+y^2}\] -\right.\right.\\
&\left.\left.- e^{2 (x+a) \sqrt{1+y^2}} \mathrm{Ei}\[-2 (x+a) \sqrt{1+y^2}\]\) \].
\end{aligned}
\eeq
In (\ref{4.21}), in the first term the integration over $dx$ is performed, while in the second term it is also possible, but the result contains an additional integration, and so in what follows it will be employed in the limit  $a \to 0$ only.

Let us start with the asymptotics for $a \to 0$. For these purposes, $\l_1$ should be  divided in following parts:
 \beq
\label{4.22}
\begin{aligned}
\l_1&= \frac{2}{\pi^2}\int\limits_{0}^{+\infty} \! d q\, \[ \sin(q a)\(\frac{\pi}{2}-\mathrm{Si}(q a) \)-\cos(q a)\mathrm{Ci}(q a)\]^2 - \\
&- \frac{2}{\pi^2}\int\limits_{0}^{+\infty} \! d q\, \[\g_E + \ln (q a) \]^2 \   \frac{4}{q\sqrt{q^2+4}}\mathrm{arcsinh}\(\frac{q}{2}\) - \\
&- \frac{2}{\pi^2}\int\limits_{0}^{+\infty} \! d q\, \( \[ \sin(q a)\(\frac{\pi}{2}-\mathrm{Si}(q a) \)-\cos(q a)\mathrm{Ci}(q a)\]^2- \right.\\
&\left.-\[\g_E + \ln (q a) \]^2\)\frac{4}{q\sqrt{q^2+4}}\mathrm{arcsinh}\(\frac{q}{2}\) \ ,
\end{aligned}
\eeq
with $\g_E$ being the Euler's constant.

The first term in (\ref{4.22}) equals to  $1/\pi a$. Here it should be noted that the integral
\beq\nonumber
\int\limits_{0}^{+\infty} \! dx\, \[ \sin(x)\(\frac{\pi}{2}-\mathrm{Si}(x) \)-\cos(x)\mathrm{Ci}(x)\]^2   \eeq
cannot be evaluated analytically (at least the authors did not succeed in searching for the answer in open sources), but the numerical calculation unambiguously indicates that its value is $\pi/2$.

The second term in $\l_1$ is explored by taking account of
\beq\nonumber
\int\limits_{0}^{+\infty} \! d q\, \frac{4}{q\sqrt{q^2+4}}\mathrm{arcsinh}\(\frac{q}{2}\) ={ \pi^2 \over 2} \ ,  \eeq
thereafter it can be rewritten as
 \beq
\label{4.23}
\begin{aligned}
&-\[ \ln^2 a + 2 \ln a \(\g_E +\frac{2}{\pi^2}\int\limits_{0}^{+\infty} \! d q\, \frac{4 \ln q}{q\sqrt{q^2+4}}\mathrm{arcsinh}\(\frac{q}{2}\) \) + \right.\\ 
&\left. + \g_E^2 + 2 \g_E \frac{2}{\pi^2}\int\limits_{0}^{+\infty} \! d q\, \frac{4 \ln q}{q\sqrt{q^2+4}}\mathrm{arcsinh}\(\frac{q}{2}\) + \frac{2}{\pi^2}\int\limits_{0}^{+\infty} \! d q\, \frac{4 \ln^2 q}{q\sqrt{q^2+4}}\mathrm{arcsinh}\(\frac{q}{2}\)\] \ .
\end{aligned}
\eeq
Again, the integrals over $d q$, entering (\ref{4.23}), cannot be found analytically, but the numerical calculation says that they are equal to $2 \ln 2$ and $\( 2 \ln 2\)^2 +\pi^2/3$, correspondingly. So (\ref{4.23}) reduces to
\beq
\label{4.24}
- \[ \( \ln a + \g_E + 2 \ln 2 \)^2 +\pi^2/3 \] \ .
\eeq
In the next step, it could be easily verified that the last term in $\l_1$ vanishes for $a \to 0$. So the nonvanishing terms in the  asymptotics of $\l_1$ for $a \to 0$ can be represented as
\beq
\label{4.25}
\l_1(a \to 0)= { 1 \over \pi a} - \( \ln a + \g_E + 2 \ln 2 \)^2 - {\pi^2 \over 3} \ .
\eeq

Now let us consider $\l_2 (a \to 0)$. For these purposes in the first integral in (\ref{4.21}), the integration region  should be divided into two pieces $(0 \, , 1/2a)$ and $(1/2a \, , \inf)$. In the first region $(0 \, , 1/2a)$, the arguments of exponents does not exceed 1, thence, upon expanding the exponents into power series and extracting from $\mathrm{Ei}$ the leading terms in $\ln a + \g_E$, one finds that the main contribution to the integral $ \int_0^{1/2a}\! dy$  comes from the latter terms only, while the others vanish for $a \to 0$.  Thereafter the integral $ \int_0^{1/2a}\! dy$ can be calculated analytically and for $a \to 0$ equals to
\beq
\label{4.26}
 \( \ln a + \g_E + 2 \ln 2 \)^2 +\pi^2/12  \ .
\eeq
In the second region  $(1/2a \, , \inf)$ for $a \to 0$, the integration is performed over asymptotically large $y$ only, therefore the integral $ \int_{1/2a}^\inf \! dy$ can be evaluated as
\beq\label{4.27}
 -\frac{2}{\pi}\int\limits_{1/2a}^{+\infty} {dy \over y^2} \ e^{2 a y}\, \mathrm{Ei}\(-2 a y\)\, \int \limits_0^\inf \! dt \ { t \cos t \over t^2 + 4 a^2 y^2}=  -\frac{4 a}{\pi}\int\limits_{1}^{+\infty} {d\x \over \x^2} \ e^\x\, \mathrm{Ei}(-\x)\, \int \limits_0^\inf \! dt \ { t \cos t \over t^2 + \x^2}
  \ .
\eeq
The integral over $d \x $ converges, and so the contribution from the second region  $(1/2a \, , \inf)$ behaves $\sim a $ for $a \to 0$.

Evaluation of the second integral in $\l_2$ for $a \to 0$ gives
\beq\label{4.28}
\frac{2}{\pi}\int\limits_{0}^{+\infty} {dy \over 1+y^2} \ \int \limits_0^\inf \! dt \ { \sin t  \over t } \arctg \( { t \over 2 a \sqrt{1+y^2} }\) \to  {\pi^2 \over 4} \ , \quad a \to 0 \ .
\eeq
As a result, the nonvanishing terms in $\l_2$ for $a \to 0$ are
\beq
\label{4.29}
\l_2(a \to 0)=\( \ln a + \g_E + 2 \ln 2 \)^2 + {\pi^2 \over 3} \ .
\eeq
Finally, for $a \to 0$ the nonvanishing part of $\l$ takes the form
\beq
\label{4.30}
\l/\a^2 \to { 1 \over \pi a} - 2\( \ln a + \g_E + 2 \ln 2 \)^2 - 2{\pi^2 \over 3} \ , \quad a \to 0 \ .
\eeq
It should be noted that (\ref{4.30}) works quite well already for $a\simeq 0.01$.

\begin{figure}[h!]
\centerline{\includegraphics[scale=0.5]{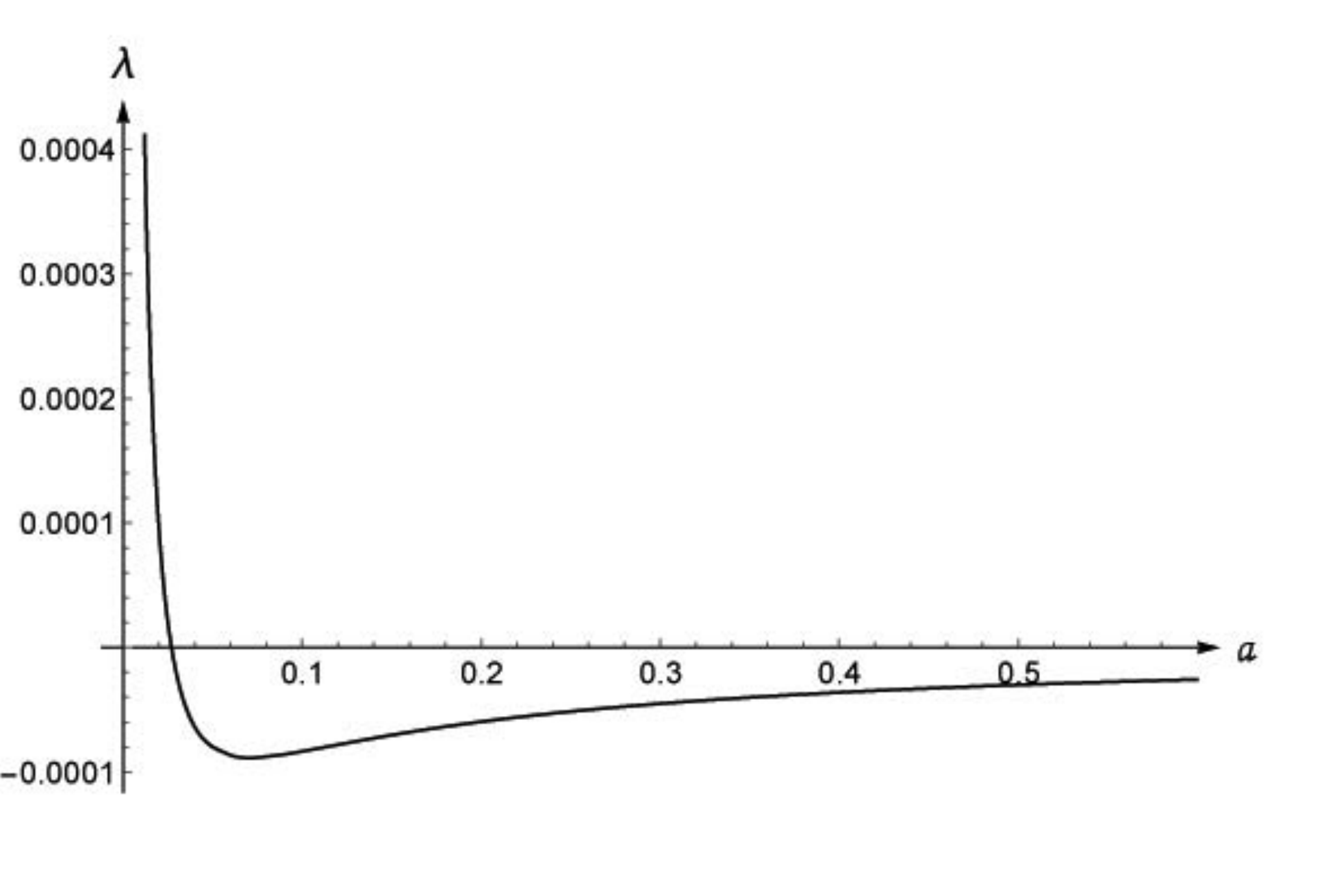}}
\caption{The renormalization coefficient $\l$ as a function of $a$. \label{fig7}}
\end{figure}

For $a \to \inf$ it could be easily found, that $\l_1 \sim 1/a^3$, while in $\l_2$ the expansion of $e^x  \mathrm{Ei}(-x)$ in the inverse powers of the argument
\beq \nonumber
e^x\,  \mathrm{Ei}(-x)=\sum\limits_{k=1} \! (-1)^k\, {(k-1)! \over x^k}
\eeq
should be used. Whence it follows that in the first integral in (\ref{4.21}), the expansion starts from $(2 a \sqrt{1+y^2})^{-3}$, and so it turns out to be $\sim 1/a^3$ also. In the second integral the expansion begins with $(2 a \sqrt{1+y^2})^{-1}$, what upon integration over $\int \! dx/(x+a)$ yields the leading term equal to $1/\pi a$. Thus, for $a \to \inf$ one obtains
 \beq
\label{4.31}
\l/\a^2 \to - { 1 \over \pi a}+ \mathrm{O}(1/a^3) \ , \quad a \to \inf \ .
\eeq
From (\ref{4.30}) and (\ref{4.31}) one immediately finds that $\l(a)$, when considered on the whole half-line $0 \leq a \leq \inf$, changes its sign and so should possess at least one zero.
The behavior of  $\l(a)$ in the intermediate region between the asymptotics  is shown in Fig.~\ref{fig7}. As expected, $\l(a)$ acquires one zero at $a_{cr}\simeq 0.027$, what corresponds to the ``size'' of the Coulomb source $\simeq 10.4$ fm.

Now let us explore the role of  $\l$ in $E_{vac}^R$. If  $\l(a_{cr})=0$, then for $a < a_{cr}$ the renormalization coefficient is positive and sufficiently large, and  there dominates in $E_{vac}^R$ the increasing $\mathrm{O}(Z^2)$-component. At the same time, for $a > a_{cr}$ the renormalization coefficient falls rapidly into negative region, and so the term $\l Z^2$ in $E_{vac}^R$ becomes a decreasing one. That is why for $Z \gg Z_{cr,1}$ there appears the effect of decreasing vacuum energy up to large negative values, since the nonrenormalized $E_{vac}$  grows in the overcritical region slower than $\mathrm{O}(Z^2)$. For $a \simeq a_{cr}$, the answer depends on  the interplay between nonperturbative effects in the phase integral, which try to decrease the vacuum energy, and the bound energy of discrete spectrum, which acts in the opposite direction. So  $E_{vac}^R(Z)$ should significantly depend on the concrete form of the external potential. In the considered case, the contribution of the discrete spectrum dominates, hence $E_{vac}^R(Z)$ increases, but it grows no longer square, but more slowly, namely  $\sim Z^\n , 1< \n <2$.

This general picture correlates completely with concrete calculations, performed for a wide range of external field parameters. In Figs.~\ref{fig8}a-d the behavior of $E_{vac}^R(Z)$ is shown for $a=0.01\, , \ a_{cr}\simeq 0.027\, , \ 0.1\, , \ 1.0 $. The number of shells for these values of $a$ is estimated as $\sim 0.006\, Z^{1.2}\, , \ 0.004 \, Z^{1.24}\, , \  0.002\, Z^{1.3}\, , \ 0.00002\, Z^{1.75}$, correspondingly. For $a=0.01$ the renormalization parameter $\l \simeq 0.0006$, hence the vacuum energy as a function of $Z$ shows up a square growth, at first due to perturbative effects, and further for $Z > Z_{cr,1}$ due to $\l Z^2$. For $a=a_{cr}$ the renormalization term disappears, and so the dominant contribution in $E_{vac}^R(Z)$ for  $Z > Z_{cr,1}$ comes from the discrete levels bound energy, which (without jumps) behaves   $\sim 0.009\, Z^{1.17}$. For $a=0.1 \ , 1.0$, on the contrary, $\l \simeq -0.00008$ and $\simeq -0.000016$ correspondingly, the nonrenormalized  $E_{vac}$ (without jumps) are estimated as $\sim 0.005\, Z^{1.2}$ and $\sim 0.0009\, Z^{1.37}$. Therefore, due to negative quadratic renormalizing term,   $E_{vac}^R(Z)$ reveals a decrease up to substantial negative values. Herewith it is clearly visible in  Figs.~\ref{fig8}c,d that there takes place first a square growth, caused by perturbative effects, which for $Z \gg Z_{cr,1}$ degrade under scenario described above, while the main contribution comes from the renormalization term $\l Z^2$ and so $E_{vac}^R(Z)$  decreases almost quadratically.
 \begin{figure} [h!]
\begin{minipage}{0.48\linewidth}
\center{ \includegraphics[scale=0.29]{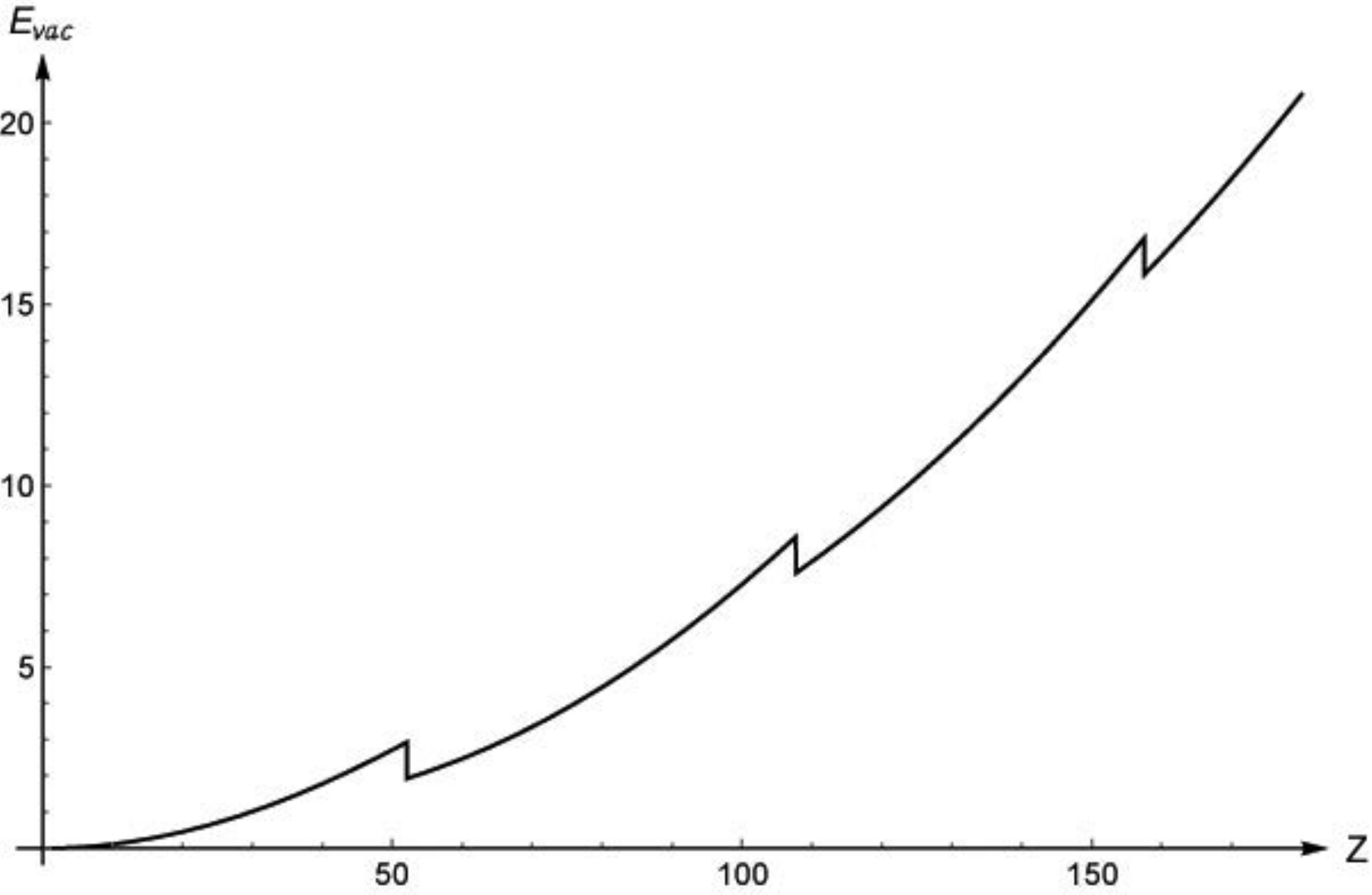} \\ a)}
\end{minipage}
\hfill
\begin{minipage}{0.48\linewidth}
 \center{ \includegraphics[scale=0.29]{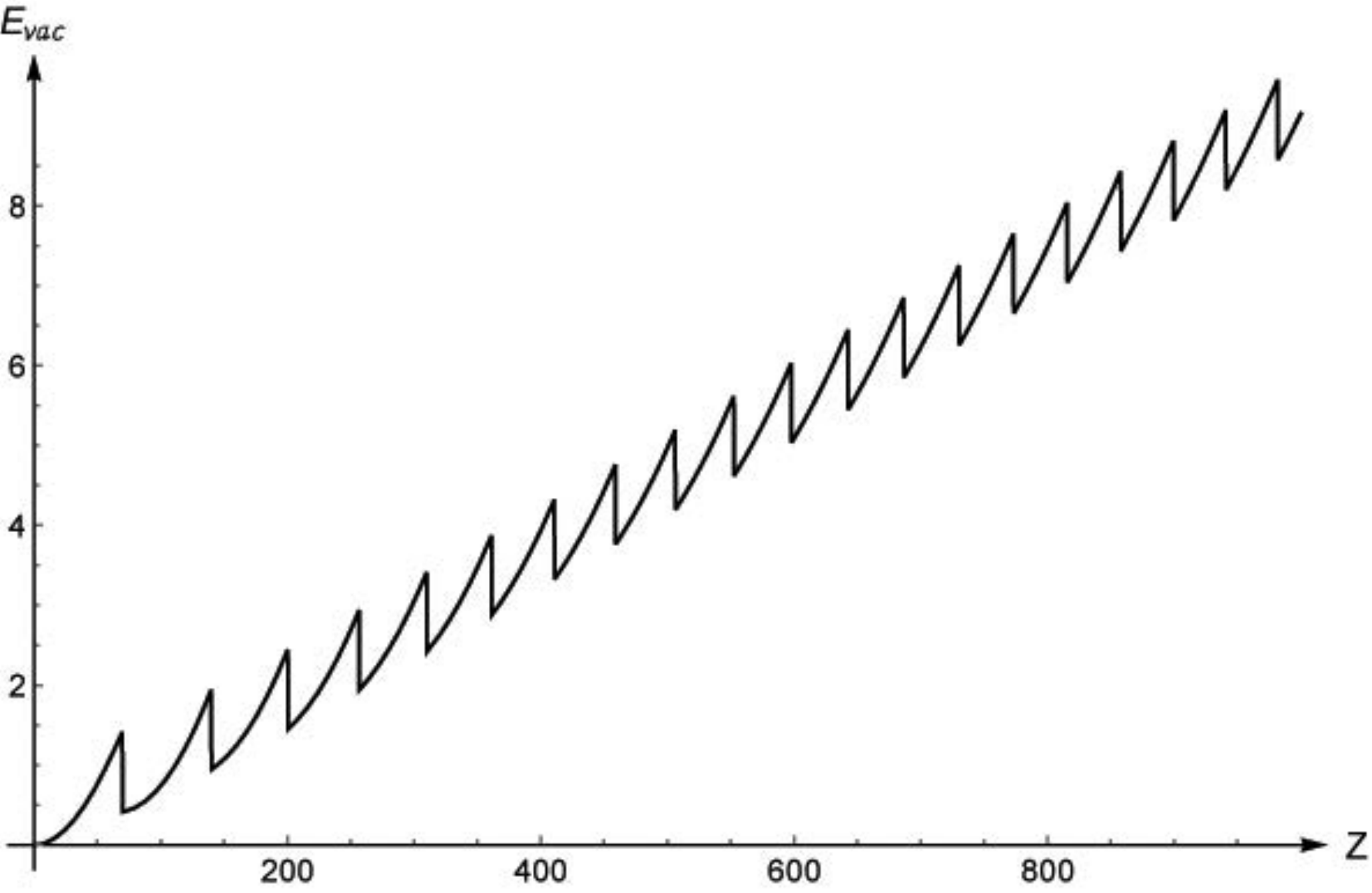}  \\ b)}
  \end{minipage}\end{figure}
 \begin{figure} [h!]
 \begin{minipage}{0.48\linewidth}
\center{ \includegraphics[scale=0.26]{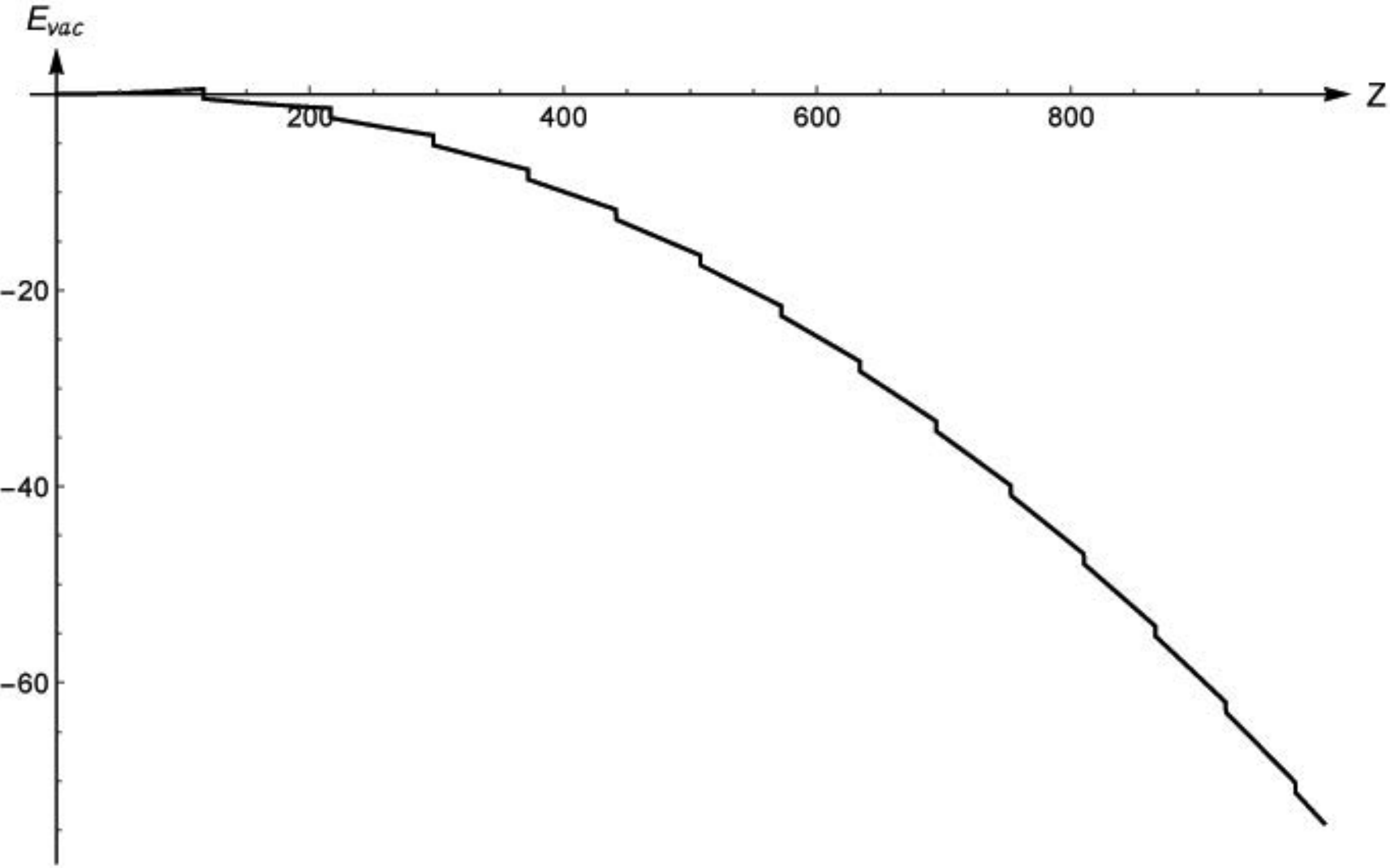} \\ c)}
\end{minipage}
\hfill
\begin{minipage}{0.48\linewidth}
 \center{ \includegraphics[scale=0.26]{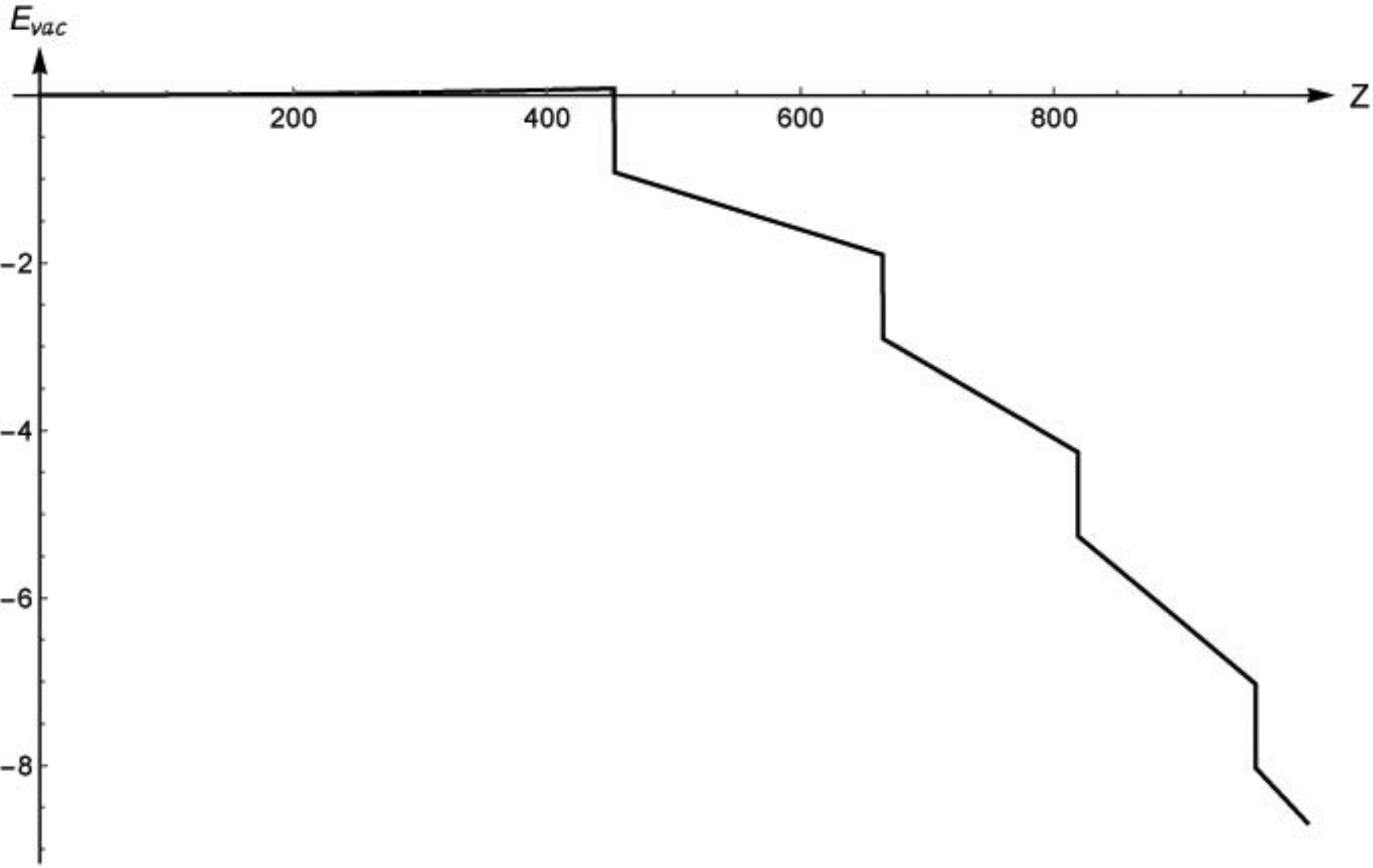}  \\ d)}
  \end{minipage}
\caption{$E_{vac}^R(Z)$  for  (a) $a=0.01$, (b) $ a=a_{cr}\simeq 0.027$, (c) $a=0.1$, (d) $a=1.0$. \label{fig8}}
\end{figure}\\
\begin{figure}[h!]
\begin{minipage}{0.48\linewidth}
\center{ \includegraphics[scale=0.25]{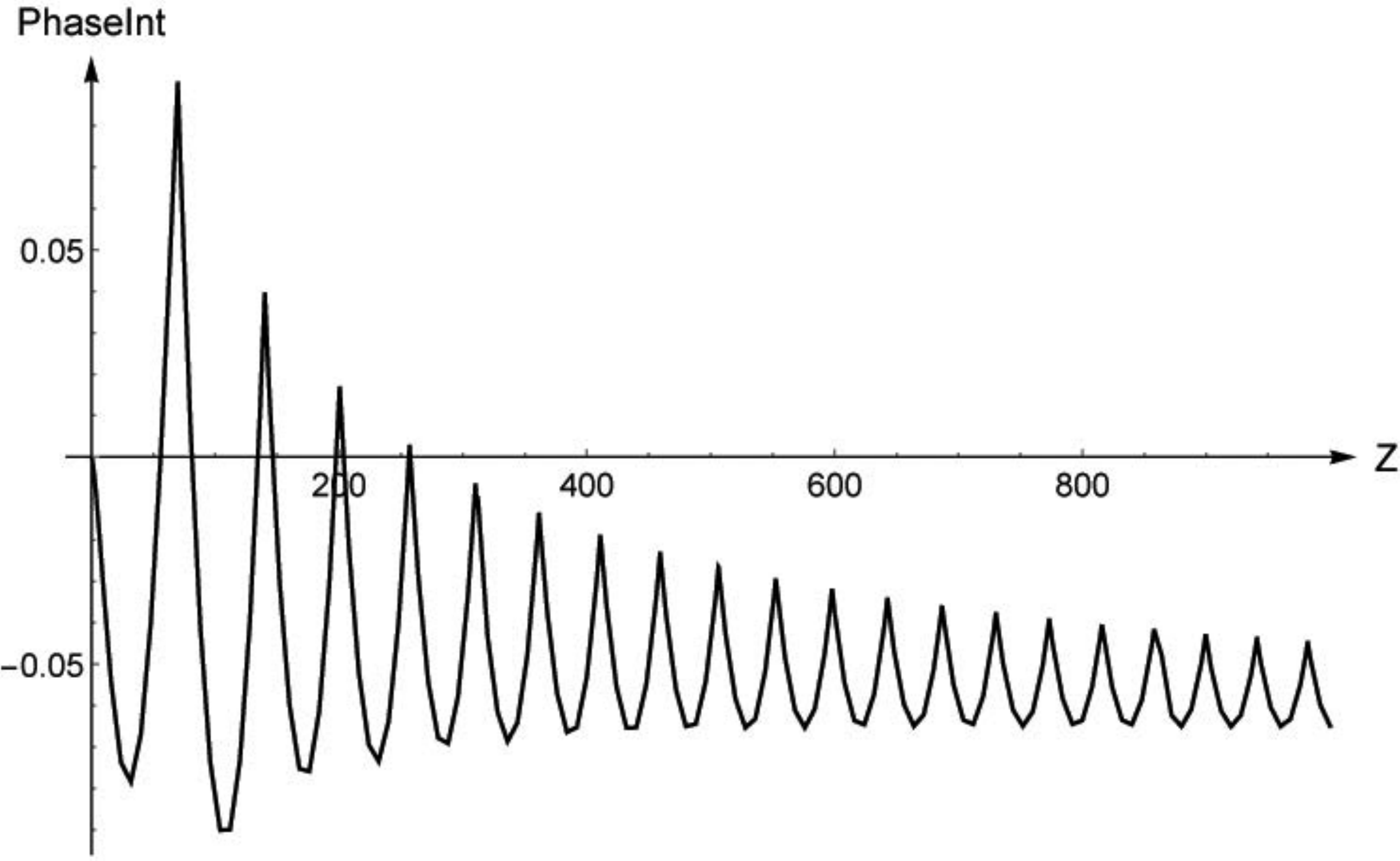} \\ a) }  \\
\end{minipage}
\hfill
\begin{minipage}{0.48\linewidth}
 \center{\includegraphics[scale=0.29]{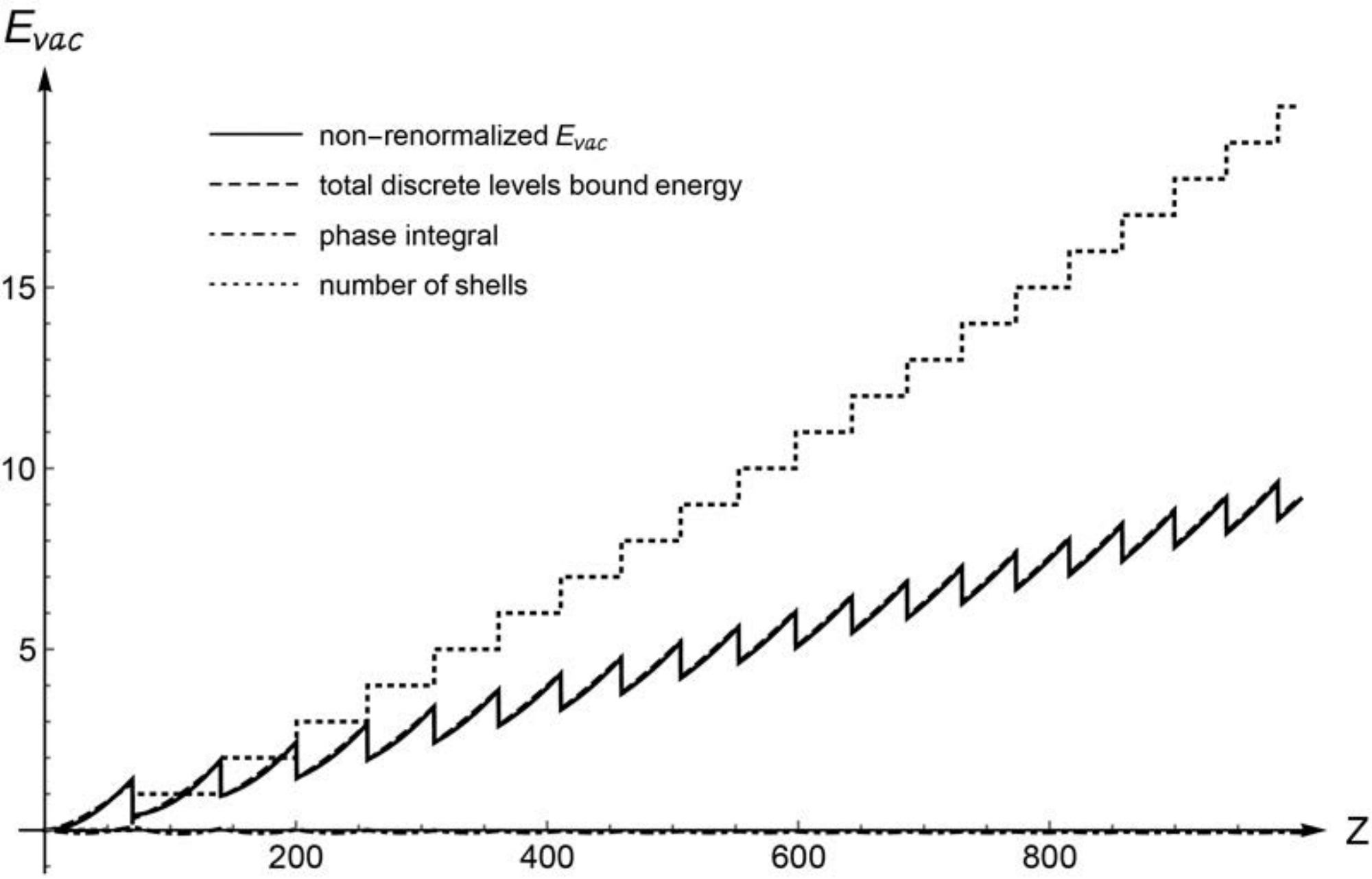} \\ b) }   \\
  \end{minipage}
\end{figure}
\begin{figure}[h!]
\center{\includegraphics[scale=0.3]{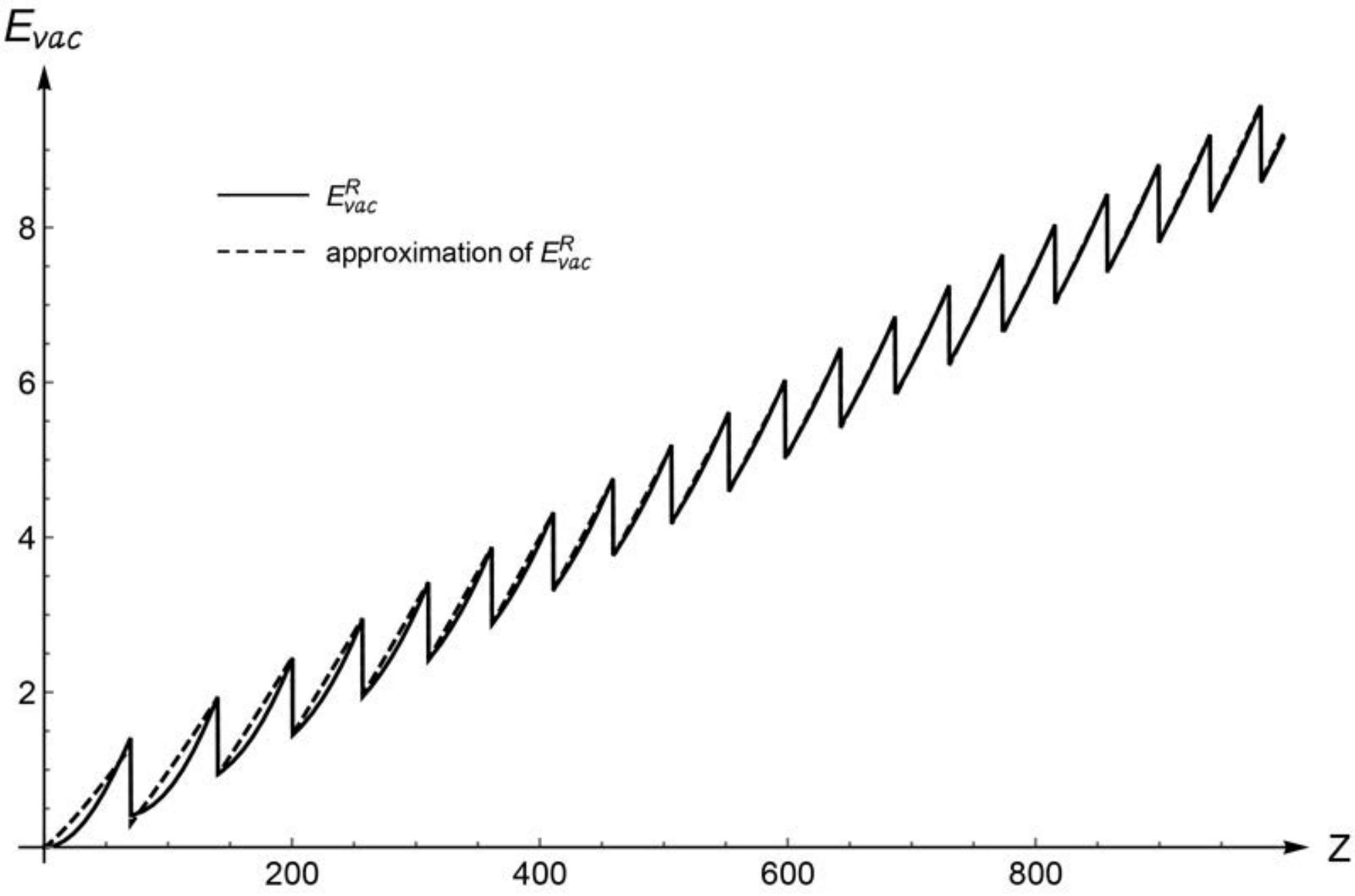} \\ c)}
\caption{(a) phase integral, (b)  nonrenormalized $E_{vac}$, total discrete levels bound energy,  phase integral and the number of shells, (c) $E_{vac}^R(Z)$ and its approximation for $a=a_{cr}$.\label{fig9}}
\end{figure}\\
\begin{figure}[h!]
\begin{minipage}{0.48\linewidth}
\center{ \includegraphics[scale=0.25]{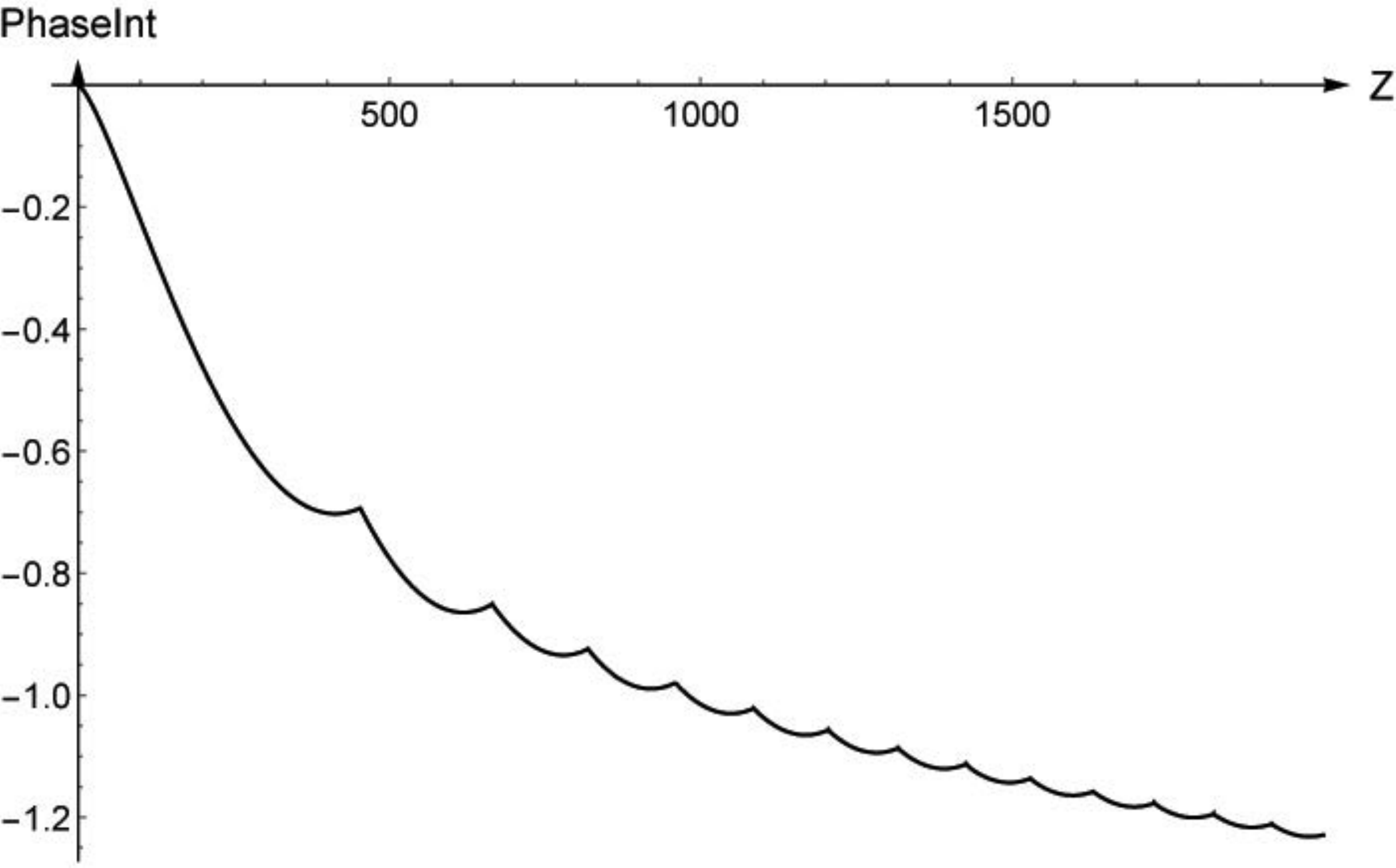} \\ a) }  \\
\end{minipage}
\hfill
\begin{minipage}{0.48\linewidth}
 \center{\includegraphics[scale=0.29]{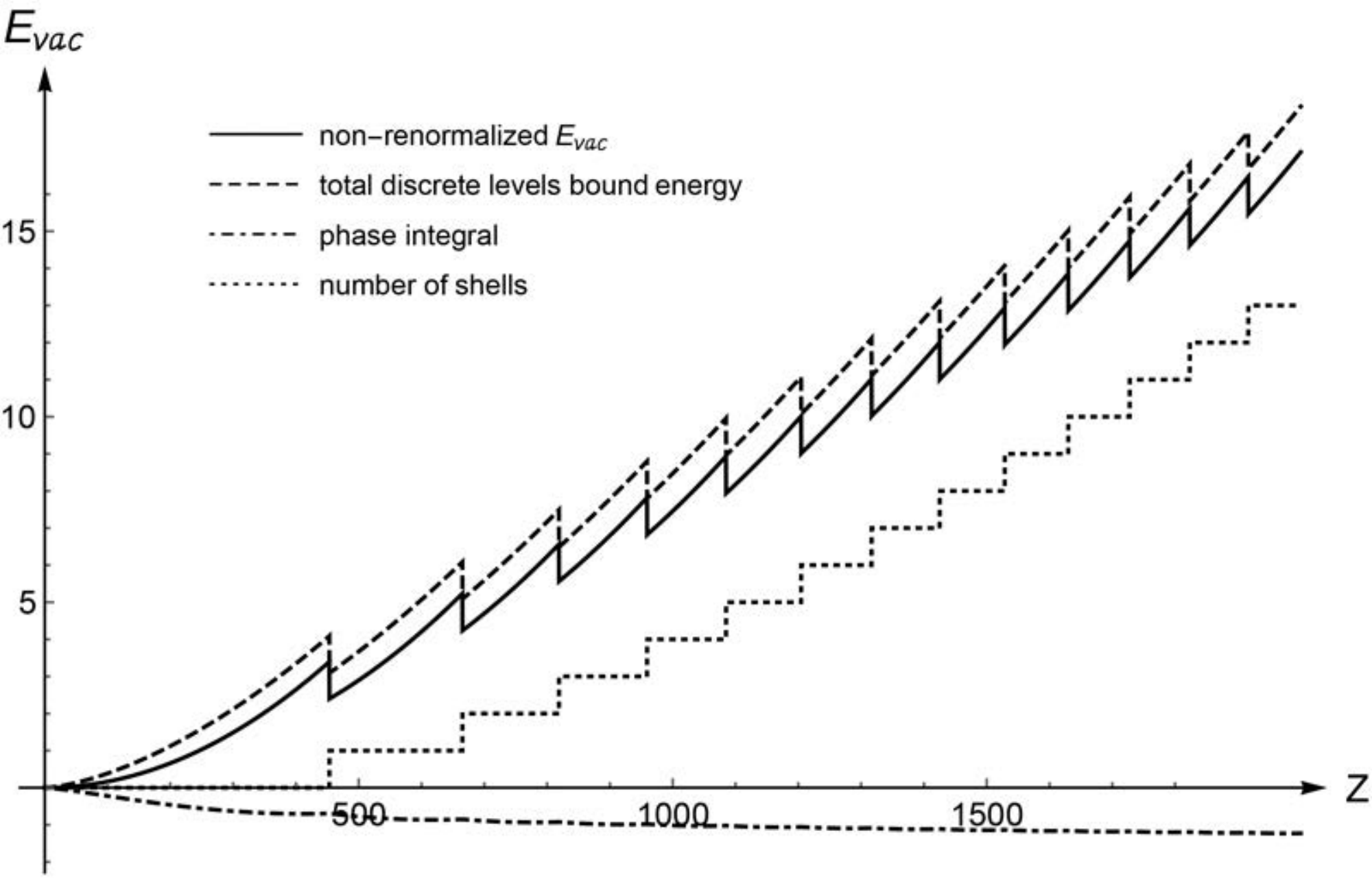} \\ b) }   \\
  \end{minipage}
\end{figure}
\begin{figure}[h!]
\center{\includegraphics[scale=0.3]{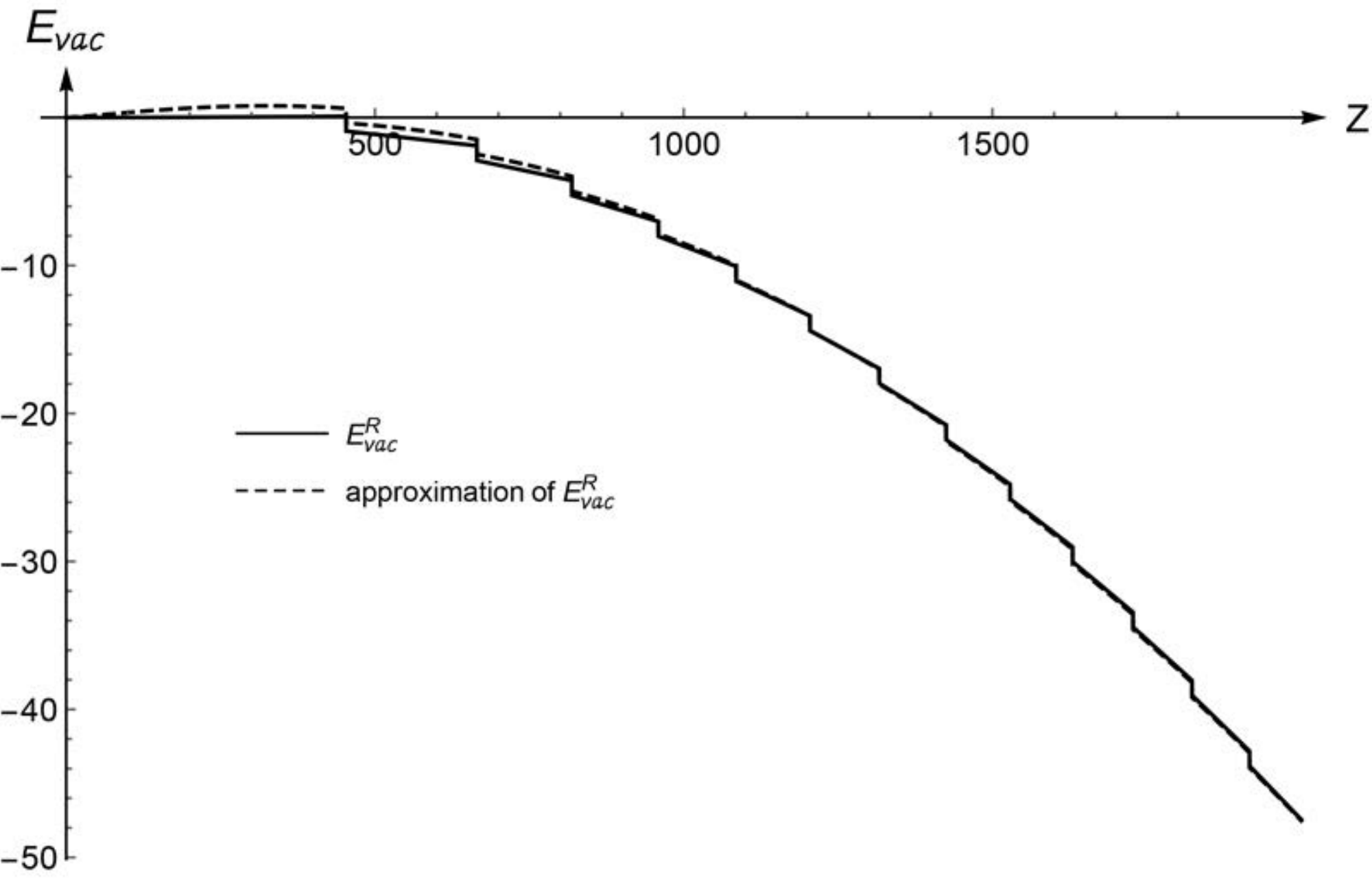} \\ c)}
\caption{(a)  phase integral, (b)   nonrenormalized $E_{vac}$, total discrete levels bound energy,  phase integral and the number of shells, (c)  $E_{vac}^R(Z)$ and its approximation for $a=1$. \label{fig10}}
\end{figure}

With more details, the formation of resulting curves for $E_{vac}^R(Z)$ looks as follows. The most representative cases here turn out to be $a=a_{cr}$ and $a=1$. For the first one, the behavior of ingredients of the final answer is shown in Fig.~\ref{fig9}. Fig.~\ref{fig9}a shows the behavior of the phase integral. As expected, the latter is an oscillating function of $Z$ with negative jumps of the derivative at $Z_{cr}$.
Fig.~\ref{fig9}b includes the nonrenormalized $E_{vac}$ (solid line), the total bound energy (dashed), the phase integral (dash-dotted) and the number of shells (dotted). As it was mentioned above, the magnitude of phase integral is much  smaller compared to the total discrete levels bound energy, and so the curves of the latter and  $E_{vac}$ are almost indistinguishable. Fig.~\ref{fig9}c shows the renormalized $E_{vac}^R(Z)$ (solid line) and its approximation (dashed), formed by the function  $0.009026\, Z^{1.17}$, which simulates the nonrenormalized $E_{vac}$ without jumps at $Z_{cr}$, and jumps  by (-1) at $Z_{cr}$. Figs.~\ref{fig10}a,b,c repeats the same details of $E_{vac}^R$ formation for $a=1$. The main difference here is that the approximation of $E_{vac}^R(Z)$ is formed now by the function $0.000903\, Z^{1.37}$ plus renormalization term $\l Z^2$ and jumps at $Z_{cr}$. Note also that the behavior of the number of shells  is quite different. For $a=a_{cr}$ the first $Z_{cr,1}$ is much less, than for $a=1$, but for $Z \gg Z_{cr,1}$ the number of shells increases for $a=1$ more rapidly compared to $a=a_{cr}$.

\begin{figure}[h!]
\begin{minipage}{0.48\linewidth}
\center{ \includegraphics[scale=0.26]{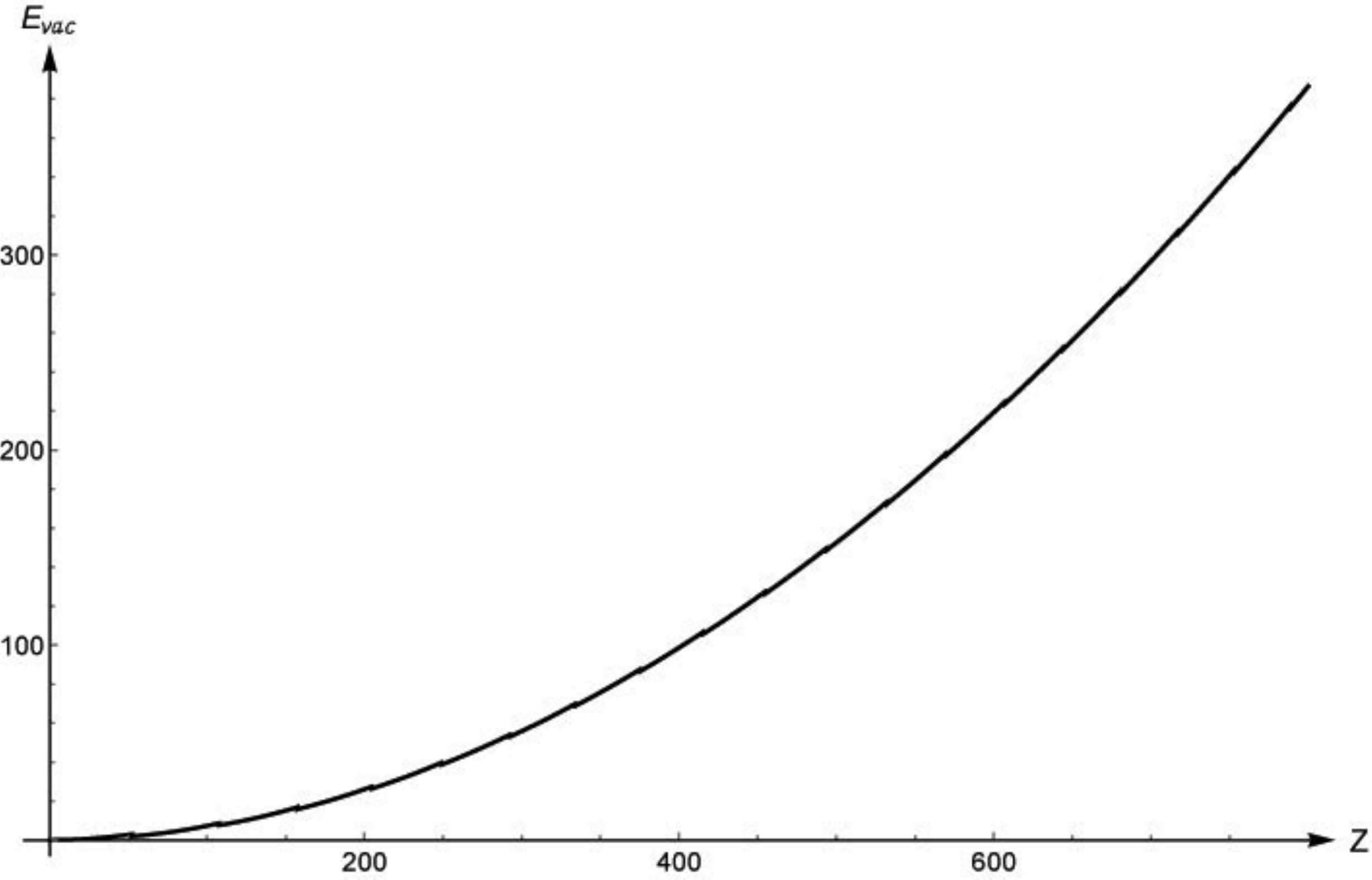} \\ a) }  \\
\end{minipage}
\hfill
\begin{minipage}{0.48\linewidth}
 \center{\includegraphics[scale=0.26]{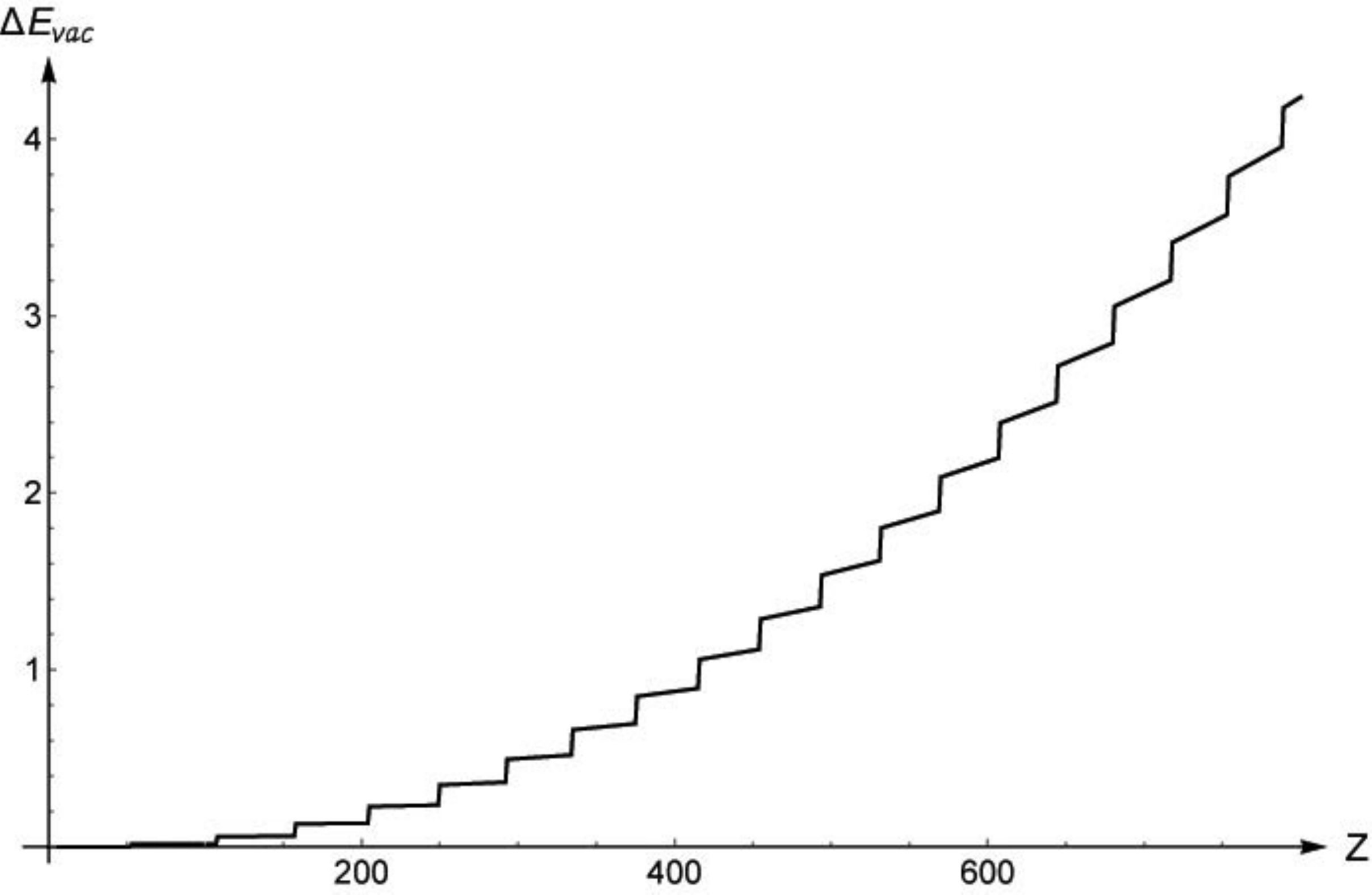} \\ b) }   \\
  \end{minipage}
  \vfill
 \begin{minipage}{0.48\linewidth}
\center{ \includegraphics[scale=0.26]{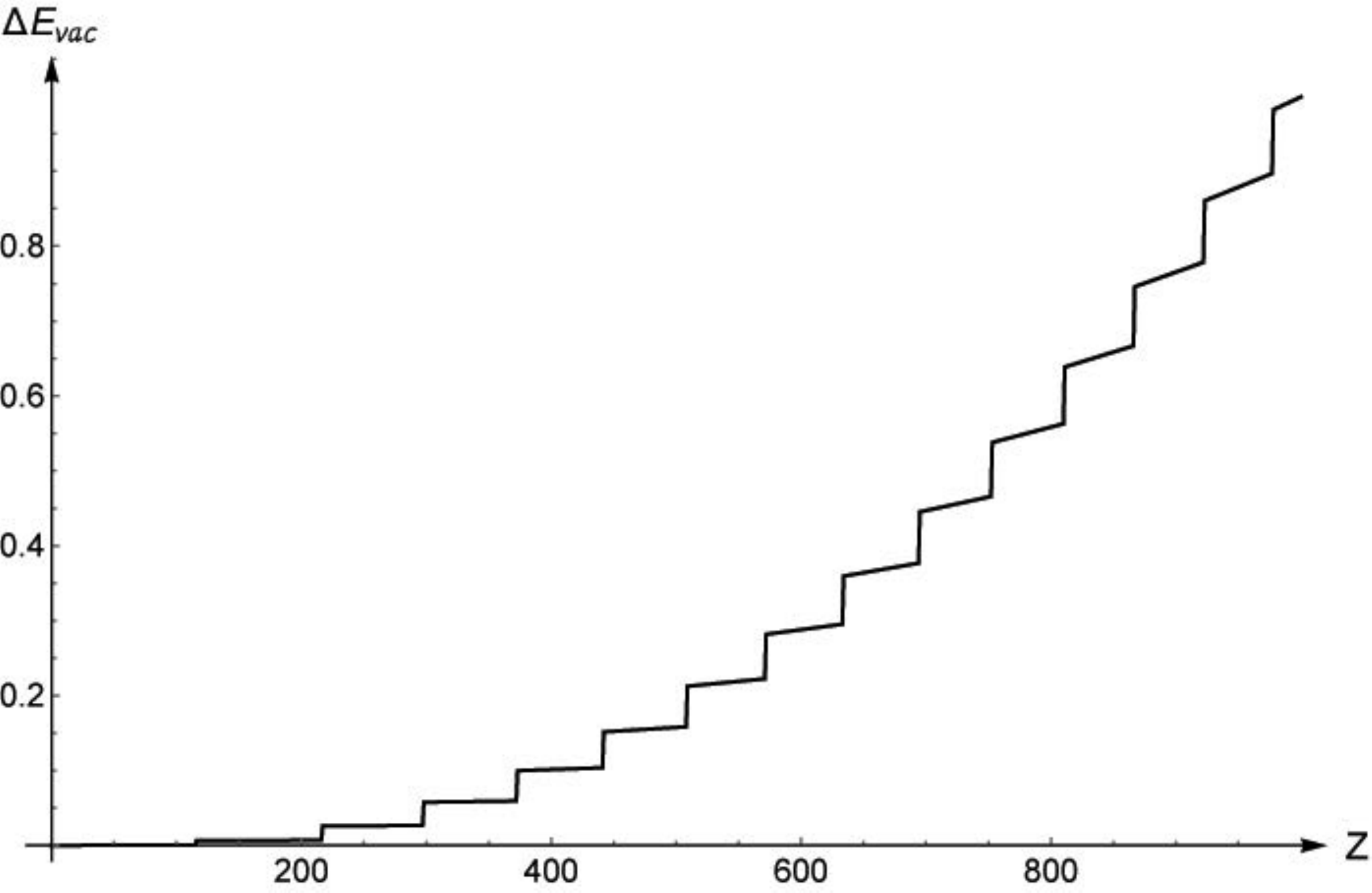} \\ c) }  \\
\end{minipage}
\hfill\begin{minipage}{0.48\linewidth}
\center{\includegraphics[scale=0.26]{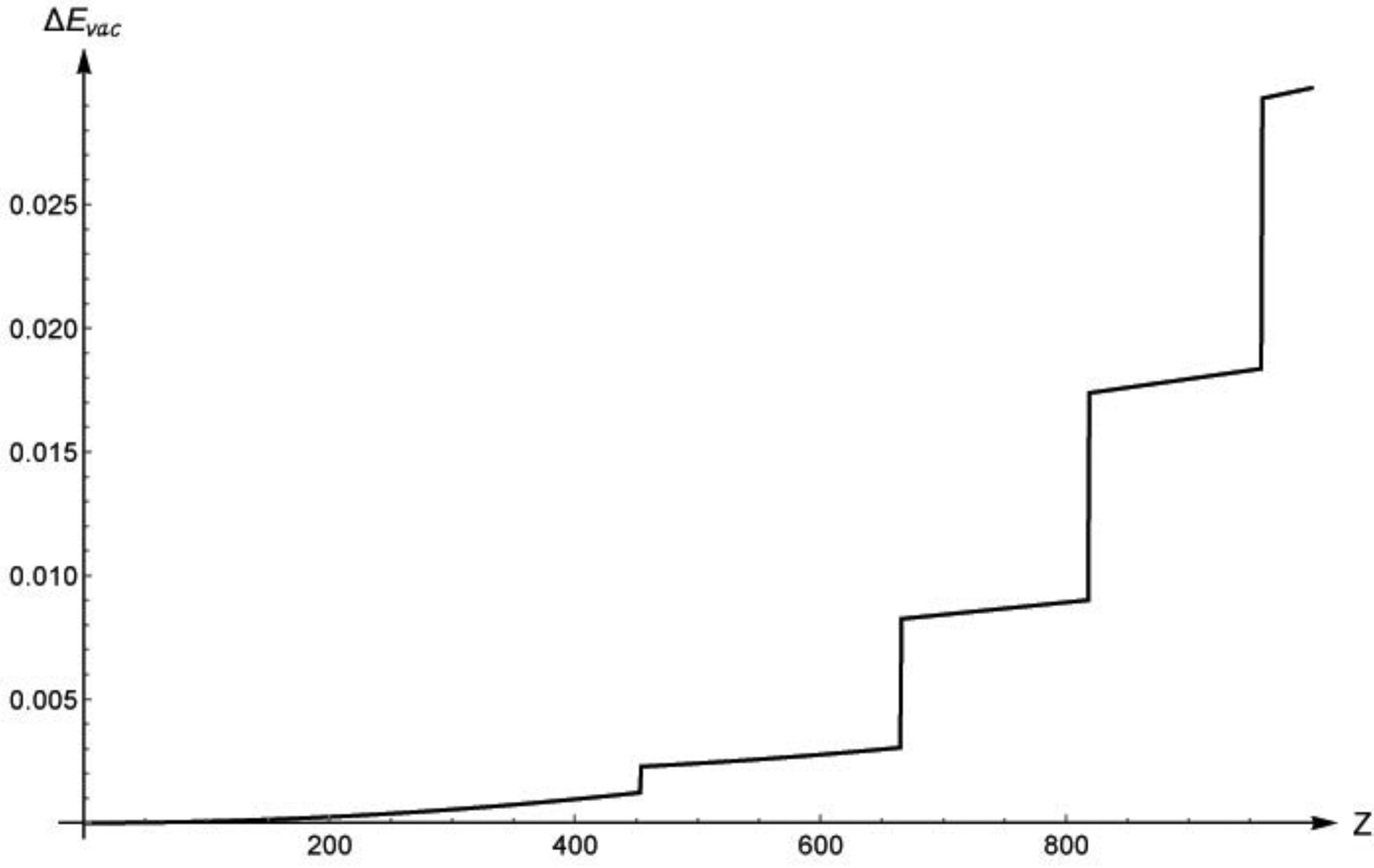} \\ d) }   \\
  \end{minipage}
\caption{(a,b)  $E_{vac}^R(Z)$ and $\D E_{vac}^R(Z)$ for $a=0.01$, (c,d)  $\D E_{vac}^R(Z)$ for $a=0.1\, , 1.0$ \label{fig11} }
\end{figure}

\section{Conclusion}

To conclude let us firstly mention that the calculation of  vacuum energy could be made completely self-consistent by means of (\ref{4.4})--(\ref{4.6}) combined with the lowest-order diagram renormalization without applying to the vacuum density and effects of shells.  However, in fact the decrease of  $E_{vac}^R$ in the overcritical region is caused indeed by the nonperturbative changes in the vacuum density for $Z > Z_{cr,1}$ due to discrete levels diving into lower continuum.  In  1+1 D, due to specifics of one-dimensional DC problem, the growth of vacuum shells number $\sim Z^s \ , \ 1<s<2$, at least for the considered range of external sources. Therefore, they are able only to decrease the speed of growth of nonrenormalized  $E_{vac}$  in the overcritical region up to $\sim Z^\n$, $1<\n<2$, and so the dominant contribution comes from the renormalization term $\l Z^2$. In more spatial dimensions, the shell effect turns out to be much more pronounced. Such behavior of $E_{vac}^R(Z)$ in the overcritical region confirms the assumption of the neutral vacuum transmutation  into the charged one under such conditions  \cite{9,10,11,13,25}, and thereby of spontaneous positron emission,  accompanying the emergence of the next vacuum shell due to total charge conservation.

It would be worthwhile to note that as in other works on vacuum polarization under strong Coulomb field \cite{21,22,23,24}, here the contribution of virtual photons was ignored, and so    only the one-loop diagram  (Fig.~\ref{fig1}) has been used for vacuum energy evaluation. A simple estimate of the effects, coming from virtual photons exchange, could be based on Coulomb energy, associated with   the vacuum charge density, with the same cutoff, as in the  external field (\ref{0.1})
\beq\label{5.1}
\D E_{vac}^R=\1/2 \int \! dx \, dx' \ {\r_{vac}^R(x)\, \r_{vac}^R(x') \over |x-x'|+ a} \ .
\eeq
Evaluation of  $\D E_{vac}^R$ shows that  at this assessment of one-photon exchange contribution to the vacuum energy  the effect turns out to be sufficiently small, about two orders less than that from the fermion loop (see Fig.~\ref{fig11}). Figs.~\ref{fig11}a,b shows  $\D E_{vac}^R$ as a function of $Z$  in comparison with $E^{R}_{vac}(Z)$, found via (\ref{4.4})--(\ref{4.6}), for $a=0.1$. Figs.~\ref{fig11}c,d demonstrate  $\D E_{vac}^R$ as a function of $Z$  for $a=0.1 \ , 1.0$. $E^{R}_{vac}(Z)$ for these values of the cutoff are already given on Fig.~\ref{fig8}c,d. Note also that in  vacuum effects, caused by virtual photons, the shell effect  shows up quite clearly.

Thus, in 1+1 D  vacuum polarization  effects could yield for pertinent parameters of Coulomb sources  such behavior of the vacuum energy in the overcritical region that is significantly different from the perturbative one. The specifics of 1+1 D shows up here in that the decrease rate of $E_{vac}^R (Z)$  in the considered range of Coulomb sources does not exceed $-|\l| Z^2$, due to extremely slow increase of vacuum shells number in one-dimensional case. Herewith, for obvious reasons, we omit the question of to what extent such a supercritical region could be physically realizable (as well as the whole one-dimensional picture for a relativistic H-like atom itself).

\bibliographystyle{ieeetr}
\bibliography{VP1D_x+a_arXiv}

\end{document}